\documentclass[aps,prd,twocolumn,longbibliography, 10pt]{revtex4-2}
\usepackage{graphicx}
\usepackage{booktabs}
\usepackage{amsmath}
\usepackage{amsfonts}
\usepackage{physics}
\usepackage{amsthm}
\newtheorem{proposition}{Proposition}
\usepackage{url}
\usepackage[colorlinks=true, allcolors=blue]{hyperref}

\begin{document}
\title{Covariant Measures of Non-Markovianity in Curved Spacetime}
\author{Tushar Waghmare}
\email[corresponding author: ]{tusharwaghmare723@gmail.com}
\affiliation{Department of Metallurgical and Materials Engineering, Indian Institute of Technology Kharagpur, WB, India. 721302}

\begin{abstract}
Standard diagnostics of quantum non-Markovianity are commonly formulated in terms of dynamical maps defined on a preferred time foliation. This becomes conceptually and operationally ambiguous in curved spacetime, where no global time coordinate exists and causal structure is primary. We develop a covariant framework for open quantum dynamics along arbitrary timelike worldlines by constructing multi-time processes (process tensors) from overlapping causal diamonds. For an Unruh--DeWitt detector weakly coupled to a scalar field in a Hadamard state, we quantify memory by the operational distance between the resulting process tensor and the convex set of Markovian (CP-divisible) combs, yielding a foliation-independent measure of non-Markovianity. Numerical benchmarks in $(1{+}1)$ dimensions compare inertial and uniformly accelerated motion, as well as static and infalling trajectories in Schwarzschild spacetime. Inertial motion is nearly Markovian, whereas acceleration, curvature, and horizons generate long-range temporal correlations and strong multi-time memory, including an activation gap: effects that remain weak in single-step diagnostics become detectable—and can be enhanced—by genuinely multi-time protocols. Our results provide an operational route to quantifying quantum memory in relativistic settings and identify acceleration, curvature, and horizons as controllable ingredients for relativistic quantum-information tasks.
\end{abstract}

\maketitle
\flushbottom

\section{Introduction}

The interface of quantum information and gravitation has sharpened concrete questions about what information is operationally accessible in quantum field theory on curved spacetime, from horizon physics to the structure of entanglement and correlations in relativistic settings \cite{Hawking1975,Unruh1976,Witten2018}. A particularly direct operational viewpoint is provided by local probes, particle detectors or observers coupled to a quantum field, whose statistics encode the field's correlations along a worldline. Because the field can store information about a probe's earlier interactions and influence later outcomes, the effective reduced dynamics of such probes are generically non-Markovian, and this memory can matter for tasks such as discrimination, control, and metrology in relativistic environments \cite{BreuerRMP2016,LiPhysRep2018}.

In flat spacetime, memory effects can often be analyzed with familiar tools: spectral densities, master equations, and influence-functional techniques provide controlled routes to identifying when a Markov approximation is reliable and when it fails \cite{Breuer2002}. In curved spacetime, however, importing standard non-Markovianity measures faces a conceptual obstacle that is fundamentally geometric. Many widely used diagnostics compare dynamical maps between time slices and implicitly rely on a preferred time parameterization \cite{BLP2009,RHP2010}. In general relativity, where global foliations need not exist and causal structure is primary, such constructions risk conflating coordinate choices with physical memory. In addition, two-time criteria are not designed to capture the full hierarchy of multi-time correlations that naturally arise in quantum fields. This limitation becomes especially sharp near horizons, where redshift and backscattering can generate long-lived ``memory tails'' and revivals that challenge simple Markovian approximations \cite{LoukoSatz2008,Hodkinson2012}. Relativistic quantum-information studies such as entanglement harvesting and detector correlations in curved backgrounds and in de~Sitter also highlight that curvature and acceleration imprint nontrivially on operational correlations \cite{HariPRD2024,MayankPRD2025}.

In many detector-based treatments, whether formulated in master-equation language or through influence functionals, one often invokes a Markov or local-in-time approximation (for example, by coarse-graining to a time-local generator) to obtain a tractable evolution for the detector’s reduced state \cite{Breuer2002,BreuerRMP2016,HuPazZhang1992}. Near strong redshift, backscattering, or acceleration-induced long correlations, these approximations can fail in ways that are not reliably diagnosed by two-time criteria alone \cite{LoukoSatz2008,Hodkinson2012}. The framework developed here provides a covariant, multi-time certificate of when memory is operationally present: it quantifies the distinguishability of the physical multi-time process from the closest CP-divisible (memoryless) comb and exposes correlations that become accessible only to genuinely multi-time protocols \cite{PollockPRL2018}.

In this work we propose a manifestly covariant route to quantifying memory for worldline-local probes. Our starting point is the process-tensor (quantum-comb) description of multi-time dynamics \cite{PollockPRL2018}. For field-coupled probes, this object can be viewed as an operational discretization of the Feynman--Vernon influence functional \cite{FeynmanVernon1963}: it encodes the statistics of arbitrary sequences of interventions on the probe and therefore captures precisely the multi-time correlations that two-time diagnostics may miss. To make this construction covariant, we build the process tensor from the field's correlation functions organized into overlapping \emph{causal diamonds} along the worldline, so that the resulting multi-time process depends only on proper time and local causal structure rather than any global coordinate choice.

We focus on the canonical Unruh--DeWitt detector model coupled to a scalar field in a Hadamard state. Within the weak-coupling regime, we define non-Markovianity as an \emph{operational distinguishability} measure: the distance between the physical multi-time process and the convex set of Markovian processes, understood here as CP-divisible combs. This definition elevates ``Markovian vs.\ non-Markovian'' from an observer-dependent slicing of dynamics to a statement about which multi-time statistics can (or cannot) be reproduced by a memoryless process compatible with the same intervention structure.

Physically, the resulting ``activation gap'' reflects the fact that the field can store trajectory-dependent information in correlations that do not appear in any single reduced map, but become accessible only when one correlates interventions across multiple time steps \cite{PollockPRL2018,MilzPRXQ2021}. Operationally, this is witnessed by multi-time channel discrimination: there exist scenarios where any two-time test has near-random success probability, while a genuinely multi-time tester distinguishes the true process from its best Markovian approximation with a significantly higher bias \cite{Chiribella2008PRL,ChiribellaPRA2009}.

We then benchmark how geometry controls memory by comparing inertial and uniformly accelerated motion in Minkowski space with static and infalling trajectories in Schwarzschild spacetime (in $(1{+}1)$ dimensions). Our benchmarks are designed to answer three concrete questions. (i) Which geometric ingredients (acceleration, curvature, and horizons) produce long-lived temporal correlations along a worldline in a way that survives operational coarse-graining? (ii) To what extent can two-time diagnostics underestimate memory compared to genuinely multi-time protocols built from the full process tensor? (iii) How do these effects differ between static and infalling observers near a black hole horizon? The remainder of the paper develops the covariant construction needed to pose these questions in a foliation-independent way and evaluates them in controlled $(1{+}1)$-dimensional settings.

\section{Covariant non-Markovianity via process tensors}
\label{sec:PT_curved}

In this section we construct the multi-time object that underlies all later diagnostics: a covariant description of the detector's reduced dynamics along a timelike worldline that does not rely on any global time foliation. Conceptually, the relevant quantity for a worldline-local probe is the Feynman--Vernon influence functional restricted to the probe trajectory: it encodes how field correlations couple different portions of the worldline. The \emph{process tensor} provides an operational discretization of this influence functional \cite{PollockPRL2018,PollockPRA2018, FeynmanVernon1963,Giarmatzi2021WitnessingMemory}: it maps a sequence of interventions on the probe to observable multi-time statistics, and it is therefore sensitive to memory that two-time diagnostics can miss.

\begin{figure*}[t] 
  \centering
  \centerline{\includegraphics[width=0.9\textwidth]{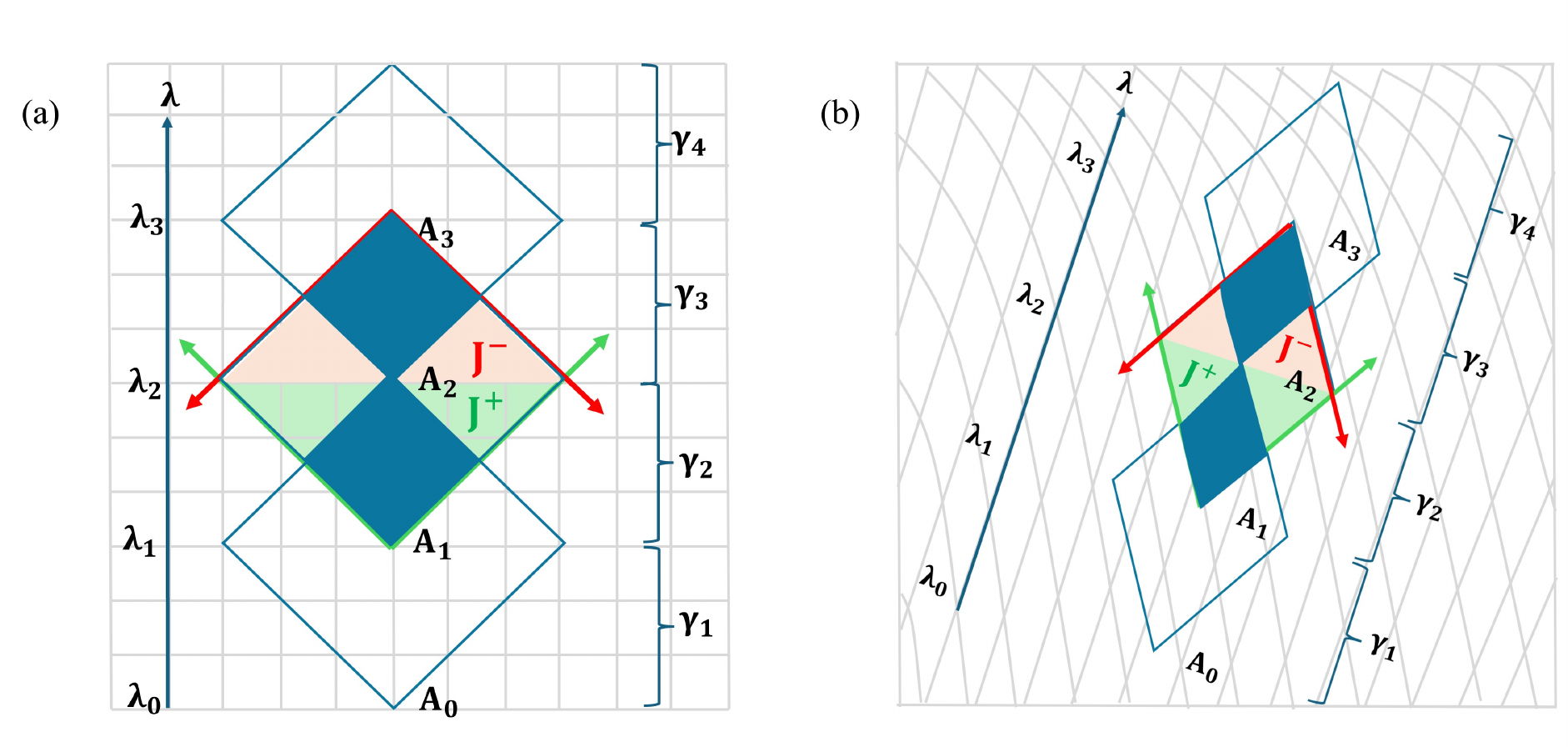}}
  \caption{{(Color online) Schematic illustration of causal diamonds and interventions along a geodesic worldline. 
(a) In flat (Minkowski) spacetime, a sequence of interventions $A_k$ is performed along a worldline parameterized by the affine parameter $\lambda$. Each segment $\gamma_k = [\lambda_{k-1},\lambda_k]$ generates a causal diamond (outlined diamonds). For the second intervention $A_2$, the light–green and light–orange cross–hatched wedges depict its causal future $J^{+}(A_2)$ and causal past $J^{-}(A_2)$, respectively; their overlap defines the shared causal volume with neighboring segments, which supports field correlations quantified by the Wightman function $W(x,x')$ and hence the memory kernels $\eta_{i-j}$. Green (red) arrows indicate future–directed (past–directed) null rays. 
(b) In curved spacetime (e.g., near an event horizon), the worldline, causal diamonds, and light cones are distorted, as suggested by the deformed background grid. Gravitational focusing and shear reduce and skew the overlap between $J^{+}$ and $J^{-}$ of neighboring interventions, leading to a smaller and asymmetric shared causal volume.}
}
  \label{fig:wide}
\end{figure*}

To make this construction manifestly covariant, we discretize the worldline into segments $\gamma_k=[\lambda_{k-1},\lambda_k]$ and group environmental contributions according to the causal diamonds generated by these segments (Fig.~\ref{fig:wide}). The resulting multi-time process depends only on (i) the causal order along the worldline and (ii) field correlation functions pulled back to the trajectory, and is invariant under coordinate reparameterizations. In the weak-coupling regime for an Unruh--DeWitt detector coupled to a scalar field in a Hadamard state, the influence of the field on the process tensor can be expressed in terms of the Wightman function and associated nonlocal memory kernels. This covariant process tensor will be the input for the distance-based non-Markovianity measure introduced in Sec.~\ref{sec:measure}.

\paragraph{Operational specification and covariance.}
An experiment in our setting is specified by (i) the spacetime and field state $(\mathcal{M},g_{\mu\nu},\rho_E)$, (ii) a timelike worldline $x(\lambda)$ as a geometric curve, (iii) a switching profile $\chi(\lambda)$ along the curve, (iv) the slot boundaries $\{\lambda_k\}$ (equivalently, the associated chain of causal diamonds), and (v) the class of admissible interventions (in the definition, arbitrary CPTP instruments on the detector, possibly with bounded ancilla). The resulting process tensor and the induced non-Markovianity measure are invariant under passive coordinate changes and orientation-preserving reparameterizations of $\lambda$, as well as under diffeomorphisms that preserve the causal-diamond support of interventions (prop.~\ref{prop:covariance}). By contrast, changing $\chi(\lambda)$ or $\{\lambda_k\}$ changes the operational task and therefore defines a different process tensor rather than a gauge-equivalent description.

\subsection{Setup and operational object}

We consider a two–level Unruh–DeWitt (UDW) detector $S$ weakly coupled to a real scalar field $\phi$ on a globally hyperbolic spacetime $(\mathcal{M},g_{\mu\nu})$~\cite{Unruh1976,LoukoSatz2006UDWClick,Schlicht2004UnruhConsiderations,LoukoSatz2008,Dappiaggi2006UnruhState}. The detector follows a timelike worldline $x(\lambda)$, parameterized by an affine parameter $\lambda$ along the trajectory. (In concrete examples we will often choose $\lambda$ to coincide with proper time $\tau$; however, retaining $\lambda$ in its general form illustrates covariance of the framework) The interaction Hamiltonian in the interaction picture for the joint system is taken to be
\begin{equation}
  H_{\mathrm{int}}(\lambda)
  = g\,\chi(\lambda)\,\sigma_x \otimes \phi\bigl(x(\lambda)\bigr),
  \label{eq:Hint}
\end{equation}
where $g \ll 1$ is a dimensionless coupling constant, $\sigma_x$ is a Pauli operator acting on the detector, and $\chi(\lambda)$ is a smooth, compactly supported switching function that localizes the interaction to a finite portion of the trajectory, as is standard in regularized detector models~\cite{Schlicht2004UnruhConsiderations,LoukoSatz2006UDWClick,LoukoSatz2008}. The field is prepared in a physically reasonable (Hadamard) state $\rho_E$ on the algebra of local observables~\cite{KayWald1991,HollandsWald2015PhysRep,FewsterVerch2013NecessityHadamard,Dappiaggi2006UnruhState}, and the initial joint state is assumed to be factorized,
\begin{equation}
  \rho_{SE}(\lambda_0) = \rho_S \otimes \rho_E.
\end{equation}
The factorized initial condition and small coupling are the standard assumptions of the Born (weak–coupling) expansion in open quantum systems~\cite{HuPazZhang1992,RivasRPP2014,BreuerRMP2016}. Both will be used explicitly below when we connect the process tensor to the Wightman function of the field.

Operationally, we imagine an experimenter who can apply a sequence of quantum operations to the detector at discrete parameter values
\begin{equation}
  \lambda_0 < \lambda_1 < \cdots < \lambda_{n-1} < \lambda_n,
\end{equation}
with $\lambda_0$ and $\lambda_n$ the initial and final parameter values of the experiment. At each intermediate point $\lambda_k$ ($k=1,\dots,n-1$) the experimenter implements a completely positive trace–preserving (CPTP) map $A_k$ on $S$; these maps can represent unitary rotations, noisy gates, measurements followed by feed–forward, or simply the identity channel~\cite{BreuerRMP2016,RivasRPP2014}. The field $E$ is not directly controlled: between intervention points, the combined system evolves unitarily under $H_{\mathrm{int}}(\lambda)$.

The \emph{process tensor} is the multi–time map that takes this entire sequence of operations as input and returns the statistics of any final measurement, or equivalently the final state of $S$ at $\lambda_n$~\cite{PollockPRA2018,PollockPRL2018,Berk2021QuantumMultiTime,Giarmatzi2021WitnessingMemory}. We write it as
\begin{equation}
  \Upsilon_{n:1} : (A_1,\dots,A_{n-1}) \mapsto \rho_S^{(\mathrm{out})},
  \label{eq:PT_definition}
\end{equation}
where $\rho_S^{(\mathrm{out})}$ is the state of the detector at $\lambda_n$ after the specified interventions. More generally, if each $A_k$ is an instrument with classical outcomes, $\Upsilon_{n:1}$ assigns joint probabilities to outcome strings~\cite{Rosset2018ResourceQuantumMemories}. From an operational point of view, $\Upsilon_{n:1}$ is the most general object that encodes how different choices of local control along the worldline lead to different observed statistics; memory effects correspond to the failure of $\rho_S^{(\mathrm{out})}$ to depend only on the \emph{most recent} intervention~\cite{PollockPRA2018,Milz2020WhenClassical,Berk2021QuantumMultiTime}.

\subsection{Choi representation and construction from the field}

To connect the abstract map [Eq.~\eqref{eq:PT_definition}] to the underlying field dynamics, it is convenient to use the Choi–Jamio{\l}kowski representation. For a single quantum channel $\mathcal{E}$ acting on a $d$–dimensional Hilbert space, the Choi operator is
\begin{equation}
  J[\mathcal{E}] = (\mathcal{E}\otimes \mathbb{I})\bigl(|\Omega\rangle\langle\Omega|\bigr),
\end{equation}
where $|\Omega\rangle = \sum_{i=1}^d |i\rangle \otimes |i\rangle$ is an (unnormalized) maximally entangled vector~\cite{Choi1975,Jamiolkowski1972}. The Choi operator encodes all action of $\mathcal{E}$ and is positive with a partial–trace constraint encoding trace preservation. In the multi–time setting one applies the same isomorphism slot by slot, obtaining a positive operator $\Upsilon$ on the tensor product of ``input'' and ``output'' Hilbert spaces associated with each time slice. This operator is often depicted as a quantum \emph{comb}: a sequence of teeth corresponding to time steps, with legs representing the in/out systems and internal links representing memory~\cite{Chiribella2008PRL,ChiribellaPRA2009,PollockPRA2018}.

Microscopically, the process tensor arises from the unitary dynamics generated by the interaction Hamiltonian. Let $U_k$ be the time–ordered unitary that evolves the joint system across the interval
\begin{equation}
  \gamma_k := [\lambda_{k-1},\lambda_k].
\end{equation}
Then
\begin{equation}
  U_k
  = \mathcal{T}\exp\!\left(
      -i \int_{\lambda_{k-1}}^{\lambda_k}
      H_{\mathrm{int}}(\lambda)\,d\lambda
    \right),
  \label{eq:Uk_def}
\end{equation}
where $\mathcal{T}$ denotes ordering in the affine parameter. The combined evolution, interspersed with the interventions $A_k$, yields
\begin{equation}
  \Upsilon(\{A_k\})
  = \mathrm{Tr}_E\!\left[
      U_n A_{n-1} U_{n-1} \cdots
      A_1 U_1 (\rho_S \otimes \rho_E)
    \right].
  \label{eq:PT_microscopic}
\end{equation}
Applying the Choi isomorphism to the multilinear map~\eqref{eq:PT_microscopic} produces a positive operator $\Upsilon$ with linear constraints that implement causality (no signalling from the future to the past along the comb)\cite{ChiribellaPRA2009,PollockPRA2018,CostaShrapnel2016QuantumCausalModelling,Oreshkov2012NoCausalOrder}. This operator is the fundamental object used in Sec.~III to define our distance–based measure of non-Markovianity.

\subsection{Environment correlations and Hadamard form}

To expose how spacetime correlations in the field generate memory, we now connect $\Upsilon$ to the two–point function of $\phi$. In the weak–coupling regime $g \ll 1$, one expands the interval unitaries $U_k$ using the Dyson series. Truncating at second order (sufficient to capture leading memory effects) gives
\begin{align}
  U_k
  &\approx \mathbb{I}
  - i \int_{\gamma_k} d\lambda\,H_{\mathrm{int}}(\lambda)
  \nonumber\\
  &\quad
  - \int_{\gamma_k} d\lambda
    \int_{\gamma_k} d\lambda'\,
    \mathcal{T}\!\left[
      H_{\mathrm{int}}(\lambda)\,
      H_{\mathrm{int}}(\lambda')
    \right]
  + O(H_{\mathrm{int}}^3),
  \label{eq:Dyson}
\end{align}
where the integrals run over the segment $\gamma_k = [\lambda_{k-1},\lambda_k]$. Inserting these expansions into Eq.~\eqref{eq:PT_microscopic} and tracing over the field replaces the time–ordered products of $H_{\mathrm{int}}$ by vacuum correlation functions of $\phi$. For a Gaussian (free) field state, Wick’s theorem reduces all $n$–point functions to products of the two–point (Wightman) function
\begin{equation}
  W(x,x') = \langle \phi(x)\,\phi(x') \rangle_{\rho_E}.
  \label{eq:Wightman_def}
\end{equation}
The pullback of $W$ to the worldline is
\begin{equation}
  C(\lambda,\lambda')
  = \langle \phi(x(\lambda))\,\phi(x(\lambda'))\rangle_{\rho_E}
  \equiv W\bigl(x(\lambda),x(\lambda')\bigr).
  \label{eq:Wightman_worldline}
\end{equation}
At second order in $g$, the entire influence of the environment on the process tensor is encoded in such two–point functions evaluated between pairs of segments along the worldline, in direct analogy with nonlocal memory kernels in quantum Brownian motion and related models~\cite{HuPazZhang1992,RivasRPP2014,BreuerRMP2016}.

In curved spacetime, physically admissible states for the field are \emph{Hadamard} states. These are characterized by the universal local structure of the Wightman function near coincidence~\cite{KayWald1991,Radzikowski1996,HollandsWald2015PhysRep,SahlmannVerch2001MicrolocalHadamard,FewsterVerch2013NecessityHadamard}. Specifically, for any Hadamard state one can write
\begin{equation}
  W(x,x')
  = \frac{u(x,x')}{\sigma(x,x') + i\epsilon}
    + v(x,x')\ln\!\bigl(\sigma(x,x') + i\epsilon\bigr)
    + w(x,x'),
  \label{eq:Hadamard}
\end{equation}
where $\sigma(x,x')$ is Synge’s world function (half the squared geodesic distance between $x$ and $x'$), $u$, $v$ and $w$ are smooth biscalars in a neighborhood of the diagonal, and $i\epsilon$ specifies the boundary condition~\cite{PoissonPoundVega2011,DecaniniFolacci2008}. (To avoid confusion, we reserve $\sigma(x,x')$ exclusively for Synge’s world function; the width of the switching functions used in the numerical benchmarks will be denoted by $w$ rather than $\sigma$.) The leading coefficient $u(x,x')$ is related to the Van Vleck–Morette determinant and encodes the universal light-cone singularity~\cite{PoissonPoundVega2011,DecaniniFolacci2008}. The biscalar $v(x,x')$ admits an expansion
\begin{equation}
  v(x,x') = \sum_{n\ge 0} v_n(x,x')\,\sigma^n,
\end{equation}
with coefficients determined recursively by local curvature invariants (DeWitt coefficients)~\cite{DecaniniFolacci2008,SahlmannVerch2001MicrolocalHadamard}. Finally, the remainder $w(x,x')$ is smooth and state-dependent; it carries global information such as thermal or horizon-induced excitations in Hartle–Hawking or Unruh states~\cite{Hawking1975,HartleHawking1976,KayWald1991,Dappiaggi2006UnruhState}.

The second–order contribution of the field to the detector dynamics can be organized into \emph{pairwise memory kernels} that couple segment $i$ to segment $j$,
\begin{equation}
  \eta_{i-j}
  = \int_{\gamma_i} d\lambda
    \int_{\gamma_j} d\lambda'\,
    \chi(\lambda)\,\chi(\lambda')\,
    W\bigl(x(\lambda),x(\lambda')\bigr).
  \label{eq:eta_def}
\end{equation}
These kernels appear as coefficients in the Born-expanded maps and, in the Choi picture, as weights multiplying system operators (cf. Sec.~III B).\cite{HuPazZhang1992,PollockPRA2018,Berk2021QuantumMultiTime} Physically, $\eta_{i-j}$ quantifies how strongly an excitation or de-excitation of the detector on segment $\gamma_j$ influences the dynamics on $\gamma_i$ through the field’s correlations. Diagonal entries $\eta_{i-i}$ encode local dissipation and decoherence, while off-diagonal entries $\eta_{i-j}$ with $i\neq j$ signal genuine temporal memory~\cite{BreuerRMP2016,RivasRPP2014}.

Using the Hadamard form~\eqref{eq:Hadamard}, the kernel naturally splits into a singular and a regular contribution,
\begin{equation}
  \eta_{i-j}
  = \eta^{\mathrm{sing}}_{i-j}
  + \eta^{\mathrm{reg}}_{i-j},
  \label{eq:eta_split}
\end{equation}
with
\begin{align}
  \eta^{\mathrm{sing}}_{i-j}
  &= \int_{\gamma_i} d\lambda
     \int_{\gamma_j} d\lambda'\,
     \chi(\lambda)\,\chi(\lambda')
  \nonumber\\
  &\quad\times
     \left[
       \frac{u\bigl(x(\lambda),x(\lambda')\bigr)}
            {\sigma(x(\lambda),x(\lambda')) + i\epsilon}
       + v\bigl(x(\lambda),x(\lambda')\bigr)
         \ln\!\bigl(\sigma + i\epsilon\bigr)
     \right],
  \label{eq:eta_sing}
\\[0.5em]
  \eta^{\mathrm{reg}}_{i-j}
  &= \int_{\gamma_i} d\lambda
     \int_{\gamma_j} d\lambda'\,
     \chi(\lambda)\,\chi(\lambda')\,
     w\bigl(x(\lambda),x(\lambda')\bigr).
  \label{eq:eta_reg}
\end{align}
The singular part $\eta^{\mathrm{sing}}_{i-j}$ is completely fixed by local geometry: it depends only on the metric through $\sigma$ and the Hadamard coefficients $u$ and $v$~\cite{PoissonPoundVega2011,DecaniniFolacci2008,HollandsWald2015PhysRep}. In contrast, $\eta^{\mathrm{reg}}_{i-j}$ carries the state–dependent, physically relevant long–range correlations that distinguish, for example, vacuum from thermal or near-horizon states~\cite{Hawking1975,HartleHawking1976}.

\section{Quantifying Covariant Non-Markovianity}
\label{sec:measure}

\begin{figure*}[t] 
  \centering
  \centerline{\includegraphics[width=0.9\textwidth]{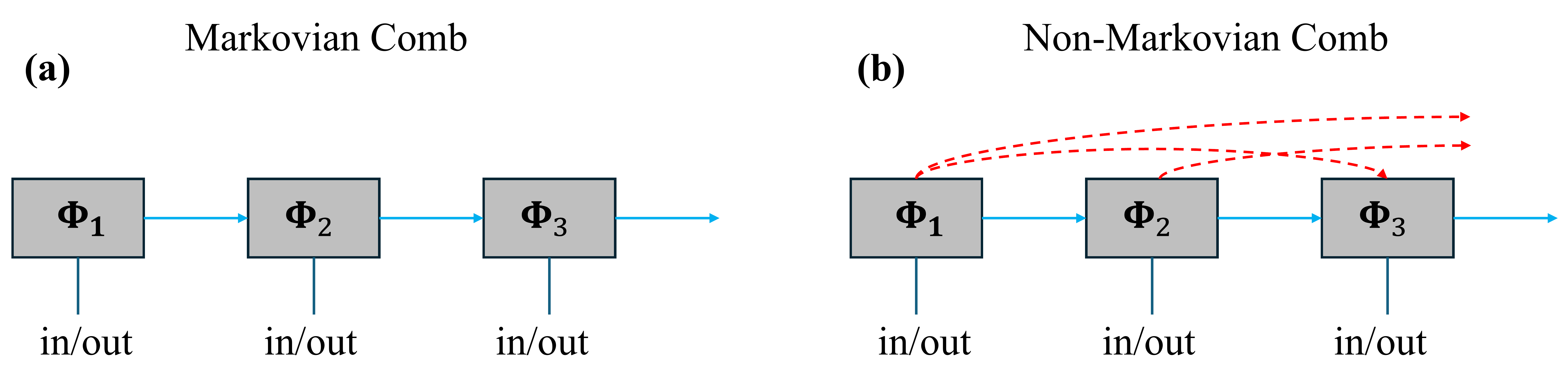}}
  \caption{Choi representation of process tensors: (a) Markovian comb, factorizing into a chain of CPTP maps $\phi_{k}$ with only nearest-neighbor causal links (blue arrows), implying memoryless dynamics. (b) Non-Markovian comb, including cross-time correlations (red dashed arrows) that capture environment-mediated memory effects, such as those from Wightman functions in curved spacetimes.}
  \label{fig:comb}
\end{figure*}

Building on the process-tensor formalism \cite{PollockPRL2018,PollockPRA2018} and a perturbative characterization of memory effects, we define an operational, diffeomorphism-invariant measure of non-Markovianity for open quantum dynamics in curved spacetimes. The measure (i) retains full multi-time correlations, including superactivation in composite probes, (ii) yields computable witnesses through Choi negativity of intermediate propagators, and (iii) remains covariant under diffeomorphisms when interventions are affinely parameterized and localized within causal diamonds. We begin by specifying the assumptions underlying our framework: the Hilbert space of the system, denoted as \(\mathcal H_S\), has a finite dimension with \(\dim\mathcal H_S=d\); time is discretized along a worldline using affine parameters \(\{\lambda_k\}_{k=0}^n\) (a finite number of intervals), and the environmental field is in a Hadamard state, allowing the Wightman function \(W(x,x')\) to be a well-defined biscalar \cite{Radzikowski1996,HW2001}. When utilizing perturbation theory, we employ the Born/weak-coupling expansion up to the second order in the coupling constants.

\subsection{Operational Definition and Resource-Theoretic Perspectives}
\label{Sec: III-A}

A process tensor \(\Upsilon\) is \emph{Markovian} if it factorizes into a chain of CPTP maps,
\begin{equation}
\Upsilon_M \quad\Longleftrightarrow\quad
\Upsilon_M = \Phi_{n:n-1}\star\Phi_{n-1:n-2}\star\cdots\star\Phi_{1:0},
\label{eq:markovian-factor}
\end{equation}
where `\(\star\)` denotes the link product that composes Choi operators into a comb \cite{ChiribellaPRA2009,BisioAPS2011}. This factorization implies that the overall dynamics can be broken down into independent, memoryless steps, where each \(\Phi_{k:k-1}\) is a completely positive and trace-preserving (CPTP) map representing the evolution from time slot \(k-1\) to \(k\). Each Choi operator $J[\Phi_{k:k-1}]$ obeys $J[\Phi_{k:k-1}] \ge 0$ and $\operatorname{Tr}_{k}\, J[\Phi_{k:k-1}] = \mathbb{I}_{k-1}$. For multi-time processes the comb-causality recursion reads
\begin{equation}
    \operatorname{Tr}_{\mathrm{out}_k}\,\Upsilon^{(k)}
= \mathbb{I}_{\mathrm{in}_k} \otimes \Upsilon^{(k-1)} \, 
\end{equation}
with ($\Upsilon^{(0)}=\mathbb{I}/d$). In such cases, CP-divisibility holds: the composite map \(\Phi_{k:j}=\Phi_{k:m}\circ\Phi_{m:j}\) is CP for all \(j<m<k\), meaning that the intermediate dynamics preserve positivity even when acting on entangled states with an ancilla \cite{RHP2010}.

To measure non-Markovianity, we adopt a distance-based approach, defining
\begin{equation}
\; N(\Upsilon) \;=\; \inf_{\Upsilon_M\in\mathcal M}\; \|\Upsilon-\Upsilon_M\|_{\rm comb} \; , \;
\label{eq:N_def_final}
\end{equation}
where \(\mathcal M\) is the convex set of all Markovian process tensors. The comb norm \(\|\cdot\|_{\rm comb}\) quantifies the operational distinguishability between two quantum processes, defined as
\[
\|X\|_{\rm comb} = \sup_{T\in\mathsf{Testers}} \|T \star X\|_1,
\]
where \(T\) ranges over causal testers (normalized multi-time instruments satisfying causality constraints), and the supremum is taken over testers with ancilla dimension bounded by \(d^{n-1}\) for \(n\) time slots \cite{ChiribellaPRA2009,Watrous2009}. We use the induced strategy norm \(\|\cdot\|_{\rm comb}\), which aligns with multi-time discrimination tasks. This norm captures the worst-case distinguishability when allowing for entanglement-assisted strategies, making it particularly suitable for quantifying deviations in quantum dynamics. Operationally, \(N(\Upsilon)\) represents the minimal distance to the nearest Markovian process, vanishing if and only if \(\Upsilon\) is Markovian. It bounds the maximal advantage an experimenter can gain in distinguishing \(\Upsilon\) from any Markovian approximation using optimal, possibly entangled, multi-time interventions.

Equivalently, by expanding the definition of the strategy norm,
\begin{equation}
    N(\Upsilon)
    = \inf_{\Upsilon_M\in\mathcal M}\;
      \sup_{T\in\mathsf{Testers}}
      \bigl\|\,T\star(\Upsilon-\Upsilon_M)\,\bigr\|_1,
    \label{eq:N_dual}
\end{equation}
where the supremum ranges over causal testers (normalized multi-time instruments obeying the usual recursion
$\operatorname{Tr}_{\mathrm{out}_k} T^{(k)} = T^{(k-1)}$ with $T^{(0)}=\mathbb{I}/d$), and the ancilla dimension can be taken no larger than the product of input dimensions up to slot $n{-}1$. Operationally, (Eq.~\ref{eq:N_dual}) is the maximal distinguishing advantage of an optimal entanglement-assisted, adaptive probing strategy between $\Upsilon$ and its closest Markovian approximation; faithfulness follows immediately since $N(\Upsilon)=0$ iff $\Upsilon\in\mathcal M$ \cite{ChiribellaPRA2009,Watrous2009}.

For completeness, we recall the link product used throughout: if $A$ and $B$ share a pair of legs labeled by $Y$, then
\begin{equation}
A\star_Y\!B\;:=\;\operatorname{Tr}_Y\!\left[(A^{T_Y}\otimes \mathbb{I}_{\overline{Y}})\,(\mathbb{I}_{\overline{Y}}\otimes B)\right],
\end{equation}
with $T_Y$ a partial transpose on the contracted space $Y$; when the contracted legs are clear, we write simply $\star$ \cite{ChiribellaPRA2009,BisioAPS2011}.

The distance $N$ defined in (Eq.~\ref{eq:N_def_final}, Eq.~\ref{eq:N_dual}) has the basic features expected of an operational resource measure for memory.
(i) \emph{Faithfulness and convexity:} $N(\Upsilon)=0$ iff $\Upsilon\in\mathcal M$, and $N$ is convex on the affine space of combs:
$N(p\Upsilon{+}(1{-}p)\Gamma) \le p\,N(\Upsilon)+(1{-}p)\,N(\Gamma)$.
(ii) \emph{Stability (Lipschitz continuity):} by the triangle inequality for the strategy norm,
$\bigl|N(\Upsilon)-N(\Gamma)\bigr|\le \|\Upsilon-\Gamma\|_{\rm comb}$,
which we use later to relate $N$ to entropic diagnostics.
(iii) \emph{Monotonicity under Markovian supermaps:} if $\mathbb{S}$ is a free transformation obtained by slot-wise pre/post-processing with CPTP channels and memoryless side channels that preserve the comb causality constraints (i.e., $\mathbb{S}[\mathcal M]\subseteq\mathcal M$), then
\begin{equation}
N\bigl(\mathbb{S}[\Upsilon]\bigr)\;\le\;N(\Upsilon),
\end{equation}
This follows from data processing for strategy discrimination and the fact that the tester set is mapped to a subset of testers under $\mathbb{S}^\dagger$ \cite{ChiribellaEPL2008,ChiribellaPRA2009,Watrous2009}.

A central feature of our construction is that it does not appeal to a foliation but to interventions localized in causal diamonds along a worldline. The following makes that precise.

\begin{proposition}[Reparametrization and diffeomorphism covariance]
\label{prop:covariance}
Let \( f \) be an orientation-preserving reparametrization of the affine parameter along the worldline and \( g \) a diffeomorphism that maps intervention diamonds to intervention diamonds (hence preserving their causal incidence). Denote by \( f_\ast \Upsilon \) and \( g_\ast \Upsilon \) the pushforwards of the process tensor, and likewise push forward testers to \( f_\ast T \), \( g_\ast T \). Then
\begin{align}
    N(f_\ast \Upsilon) &= N(\Upsilon), \\
    N(g_\ast \Upsilon) &= N(\Upsilon).
\end{align}
\textit{Proof Sketch: }The Markovian set \( \mathcal{M} \) is defined by positivity and linear trace-preservation constraints on the step-Choi operators together with the causal recursion; these constraints are form-invariant under the relabeling of slots induced by \( f \) and under the isometries induced by \( g \) on the in/out legs of each diamond. Moreover, the strategy norm is invariant under these relabelings/isometries because testers push forward to testers and \( \| f_\ast T \star f_\ast X \|_1 = \| T \star X \|_1 \) (similarly for \( g \)). Taking the infimum over \( \mathcal{M} \) gives the claim.
\end{proposition}

\emph{Resource-theoretic viewpoint.} We hence regard \emph{Markovian supermaps}—slot-local CPTP pre/post-processing with memoryless ancillary wires, diamond-preserving coarse-grainings (slot merging), and classical randomness—as free. Proposition~\ref{prop:covariance} ensures that $N$ is not only a monotone under these free operations but also invariant under reparametrizations of proper time and under diffeomorphisms that carry diamonds to diamonds, making the measure intrinsic to the causal support of the interventions. In practice, we compute $N$ via an SDP relaxation (Appendix~D) or report certified lower/upper surrogates (Sec.~IV) when the full optimization becomes costly for large $n$ \cite{Watrous2009}.

As an information–theoretic complement to the operational measure \(N(\Upsilon)\), we quantify memory through conditional mutual informations (CMIs) on the normalized comb state \(\rho_\Upsilon := \Upsilon/\Tr\Upsilon\). For an \(n\)-slot process with input/output legs \((\mathrm{in}_k,\mathrm{out}_k)\), consider the canonical cut at slot \(k\) with the tripartition
\begin{align*}
\mathsf{Past}_{\le k-1} &:= \{\mathrm{in}_0,\mathrm{out}_1,\ldots,\mathrm{out}_{k-1}\}, \\
\mathsf{Boundary}_k &:= \{\mathrm{in}_k,\mathrm{out}_k\}, \\
\mathsf{Future}_{\ge k+1} &:= \{\mathrm{in}_{k+1},\mathrm{out}_{k+1},\ldots,\mathrm{in}_n,\mathrm{out}_n\}.
\end{align*}

We define the CMI–based indicator
\begin{equation}
N_{\mathrm{ent}}^{\mathrm{CMI}}(\Upsilon)
\;:=\;
\sum_{k=2}^{n-1} I\!\left(\mathsf{Past}_{\le k-1}:\mathsf{Future}_{\ge k+1}\,\middle|\,\mathsf{Boundary}_k\right)_{\rho_\Upsilon},
\label{eq:entropic_cmi}
\end{equation}
with \(I(A:C|B) := S(AB)+S(BC)-S(B)-S(ABC)\). Intuitively, each summand measures how strongly the distant past and future remain correlated once the local interface at slot \(k\) is revealed; memoryless (CP–divisible) dynamics break those long–range correlations across every cut \cite{HaydenCMP2004,FawziRennerCMP2015}. This diagnostic has three features we rely on: \\
(i) it \emph{vanishes for CP–divisible (Markovian) combs} and, conversely, vanishing of all CMIs across the cuts, together with the standard comb–causality constraints, characterizes quantum–Markov combs that coincide with our Markovian set \(\mathcal{M}\) \cite{HaydenCMP2004,FawziRennerCMP2015}; \\
(ii) it is \emph{stable} under small perturbations of process tensor \(\Delta\Upsilon\), obeying continuity bounds of the form \(\Delta N_{\mathrm{ent}}^{\mathrm{CMI}}=O\!\big(\|\Delta\Upsilon\|_{\mathrm{comb}}\log D\big)\), where \(D\) is the relevant Hilbert–space dimension \cite{AlickiFannesCMP2004}; and \\
(iii) it is \emph{nonincreasing under the subclass of Markovian supermaps} considered in our resource theory, namely those that act locally on the in/out legs around each cut and preserve the comb causal structure, by data processing for quantum CMIs \cite{FawziRennerCMP2015}. The indicator is covariant under reparametrizations and diffeomorphisms that map diamonds to diamonds: tripartitions are carried to tripartitions, and CMIs remain invariant under the induced isometries.

We emphasize that \(N_{\mathrm{ent}}^{\mathrm{CMI}}\) is a \emph{diagnostic} rather than a replacement for \(N(\Upsilon)\): it provides an interpretable, dimension–aware proxy for long–range temporal correlations and, by the continuity bound above, scales at most linearly (up to \(\log D\) factors) with the operational distance \(N(\Upsilon)\). In practice we report \(N_{\mathrm{ent}}^{\mathrm{CMI}}\) alongside the LB/UB surrogates of Sec.~IV when computing the full strategy–norm distance is intractable.

\subsection{Witnessing and Bounding Non-Divisibility}
\label{Sec: III-B}

To witness non-Markovianity without full optimization, we extract intermediate maps $\Phi_{k:j}$ from $\Upsilon$ by contracting irrelevant legs with identity instruments (a causal tester localized to the complementary diamonds). If $\Phi_{k:j}$ is not CP for some $j < k-1$, this signals non-CP-divisibility and hence memory, as CP-divisibility requires all composite maps to preserve positivity on entangled extensions \cite{RHP2010}. The Choi operator provides a direct certificate: negative eigenvalues indicate non-CP behavior. Explicitly, 
\begin{equation}
\chi_{\Phi_{k:j}} = (\Phi_{k:j} \otimes I)(|\Omega\rangle\!\langle\Omega|),
\end{equation}
with $\operatorname{Tr}\chi_{\Phi_{k:j}} = d$ and $|\Omega\rangle = \sum_{i=1}^d |i,i\rangle$ the unnormalized maximally entangled state. The Choi-negativity, measuring the extent of negativity, is
\begin{equation}
N_\chi(\Phi_{k:j}) = \tfrac{1}{2}\bigl(\|\chi_{\Phi_{k:j}}\|_1 - d\bigr) = \sum_{\lambda_i < 0} |\lambda_i|,
\end{equation}
where $\|\cdot\|_1$ is the trace norm. A standard channel-level bound follows:
\begin{equation}
\inf_{\Phi_{\mathrm{CP}}}\ \|\Phi_{k:j} - \Phi_{\mathrm{CP}}\|_{\diamond}\ \ge\ \frac{2}{d}\,N_\chi(\Phi_{k:j}),
\label{eq:map_diamond_LB}
\end{equation}
with the infimum over CPTP maps \cite{Watrous2009}.

\paragraph*{Lifting to a comb-level bound.}
We extend Eq.~\eqref{eq:map_diamond_LB} to lower-bound the global measure $N(\Upsilon)$ from Eq.~\eqref{eq:N_def_final} (Sec.~\ref{Sec: III-A}). Consider the causal tester $T_{k:j}$ that isolates the $(j\to k)$ block with identity instruments elsewhere (normalized per comb causality). Since $T_{k:j}$ is admissible, the strategy norm satisfies
\begin{equation}
\|T_{k:j} \star (\Upsilon - \Upsilon_M)\|_1 \le \|\Upsilon - \Upsilon_M\|_{\mathrm{comb}}
\end{equation}
for any $\Upsilon_M \in \mathcal{M}$. Optimizing over $\Upsilon_M$ and applying Eq.~\eqref{eq:map_diamond_LB} to the isolated $\Phi_{k:j}$ yields the following.

\begin{proposition}[LB via intermediate Choi]
\label{prop:LB-choi-comb}
For any process tensor $\Upsilon$,
\begin{equation}
N(\Upsilon)\ \ge\ \max_{j<k}\ \frac{2}{d}\,N_\chi(\Phi_{k:j}).
\label{eq:global_LB}
\end{equation}
\end{proposition}

\begin{proof}[Proof sketch]
By definition of the strategy norm, $\|\Upsilon - \Upsilon_M\|_{\mathrm{comb}} = \sup_T \|T \star (\Upsilon - \Upsilon_M)\|_1 \ge \|T_{k:j} \star (\Upsilon - \Upsilon_M)\|_1 = \|\Phi_{k:j} - \Phi^M_{k:j}\|_{\diamond}$ for the isolated intermediate map $\Phi^M_{k:j}$ of $\Upsilon_M$. Taking the infimum over $\Upsilon_M \in \mathcal{M}$ (noting $\Phi^M_{k:j}$ is CPTP for Markovian combs) and using Eq.~\eqref{eq:map_diamond_LB} gives the bound. The max is over non-neighboring pairs to detect long-range memory. Invariance under reparametrizations and diffeomorphisms (Prop.~\ref{prop:covariance}) follows as the extraction and tester are diamond-localized.
\end{proof}

This LB is computationally efficient (via eigenvalue decomposition of $\chi_{\Phi_{k:j}}$) and certifies non-Markovianity when positive; it is tight for two-slot processes where the comb norm reduces to the diamond norm. Unlike flat-space RHP witnesses, it is covariant and captures curvature-induced effects, as benchmarked in Sec.~IV \cite{RHP2010}.

\paragraph*{Microscopic origin in the relativistic setup.}
In our UDW model [Eq.(\ref{eq:Hint})], the map is \(\Phi_{k:j}(\cdot) = \operatorname{Tr}_E[U_{k:j} (\cdot \otimes \rho_E) U_{k:j}^\dagger]\) with $U_{k:j}$ the time-ordered exponential over $\gamma_{k:j} = \bigcup_{m=j+1}^k \gamma_m$. In weak coupling ($g \ll 1$), a second-order Born expansion yields
\begin{align}
\Phi_{k:j}(\cdot) &\approx \mathbb{I}(\cdot) - i [K_{k:j}, \cdot] \notag \\
&\quad - \tfrac{1}{2} \int_{\gamma_{k:j}} \int_{\gamma_{k:j}} T \bigl[ [H_{\mathrm{int}}(\lambda), [H_{\mathrm{int}}(\lambda'), \cdot]] \bigr] d\lambda\, d\lambda' \notag \\
&\quad + O(g^3),
\end{align}
where $K_{k:j} = \int_{\gamma_{k:j}} \langle H_{\mathrm{int}}(\lambda) \rangle_E d\lambda$ (often zero) and the double commutator encodes dissipation via the Wightman function $W$. As in standard weak-coupling treatments of open quantum systems~\cite{BreuerRMP2016,RivasRPP2014}, the second-order Born expansion provides a controlled approximation to the reduced dynamics on time scales $t \lesssim O(g^{-2})$, with any violations of complete positivity signalling use beyond its nominal regime of validity. Outside this regime one would need a nonperturbative construction of $\Upsilon$, for instance via resummation schemes or tensor-network discretizations of the field, which lies beyond the scope of the present work. The Choi operator approximates
\begin{equation}
\chi_{\Phi_{k:j}} \approx \mathbb{I} + \sum_{m<l \in [j,k]} \eta_{m-l} (\sigma_x \otimes \sigma_x) + O(g^4),
\end{equation}
with pairwise kernels
\begin{equation}
\eta_{m-l} = g^2 \int_{\gamma_m} \int_{\gamma_l} \chi(\lambda) \chi(\lambda') \operatorname{Re} W\bigl(x(\lambda), x(\lambda')\bigr) d\lambda\, d\lambda'.
\end{equation}
Long-range tails in $W$ (e.g., horizon effects) drive negativity \cite{Radzikowski1996,HW2001,Unruh1976,LoukoSatz2008,PKMM2015}. For a qubit ($d=2$), eigenvalues are approximately $1 \pm \sum_{m<l} \eta_{m-l} + O(g^4)$; negativity occurs if $\sum_{m<l} \eta_{m-l} < -1$. Thus, a weak-coupling diagnostic is
\begin{equation}
N(\Upsilon) \gtrsim \max_{j<k} \biggl| \sum_{m<l \in [j,k]} \eta_{m-l} \biggr| - 1 + O(g^4),
\end{equation}
where the threshold $-1$ normalizes against the identity (derived from the qubit Choi spectrum); we use the certified Eq.~\eqref{eq:global_LB} for exact bounds in numerics.

\paragraph*{Upper bound via projection.}
For completeness, we compute an upper bound as the distance to a nearest CP-divisible comb via SDP (App.~D):
\begin{equation}
\mathrm{UB}(\Upsilon) := \min_{\Upsilon_M \in \mathcal{M}} \|\Upsilon - \Upsilon_M\|_{\mathrm{comb}},
\end{equation}
which equals $N(\Upsilon)$ but is approximated for large $n$ via relaxation. In Sec.~IV, we report UB with the LB Eq.~\eqref{eq:global_LB} to bracket $N(\Upsilon)$; both are covariant under diamond-preserving transformations \cite{Watrous2009}.

\subsection{Multi-Time Phenomena and Superactivation in composite systems}
\label{Sec: III-C}
A distinctive feature of the strategy-norm measure in Sec.~\ref{Sec: III-A} is that non-Markovianity can become \emph{operationally visible} through causal testers that coherently bridge multiple slots, akin to superactivation in quantum resources \cite{SmithYardScience2008}. Segments that appear Markovian in isolation can thus yield a nonzero comb distance $N(\Upsilon)$ when probed with entanglement-assisted, adaptive strategies across larger temporal windows.

\paragraph*{Definition (activation gap).}
Fix two non-adjacent temporal segments $S_1 = [j_1+1:k_1]$ and $S_2 = [j_2+1:k_2]$ with $k_1 < j_2$, and let $\Phi_{k_1:j_1}$ and $\Phi_{k_2:j_2}$ be their isolated intermediate maps, extracted by contracting complementary legs with identity instruments (Sec.~III~B). Define the stitched intermediate map across the union as
\begin{equation}
\Phi_{k_2:j_1}^{(S_2 \triangleleft S_1)} := \Phi_{k_2:j_2} \circ \Phi_{k_1:j_1},
\end{equation}
obtained by a causal tester that links $S_1$'s output to $S_2$'s input via an identity channel (entanglement possible if ancillas are used), with its Choi operator $\chi_{\Phi_{k_2:j_1}^{(S_2 \triangleleft S_1)}}$. The \emph{activation gap} is
\begin{equation}
\Delta_{\mathrm{act}} := \frac{2}{d} N_\chi\bigl(\Phi_{k_2:j_1}^{(S_2 \triangleleft S_1)}\bigr) - \max\left\{ \frac{2}{d} N_\chi\bigl(\Phi_{k_1:j_1}\bigr), \frac{2}{d} N_\chi\bigl(\Phi_{k_2:j_2}\bigr) \right\}.
\label{eq:activation_gap}
\end{equation}

Superactivation occurs whenever $\Delta_{\mathrm{act}} > 0$, indicating memory effects undetectable in marginal probes but revealed by coherent multi-time interventions (cross-time links in Figure~\ref{fig:comb}(b)). From the perspective of relativistic quantum information, these multi-probe effects are complementary to entanglement-harvesting protocols in which UDW pairs act as curvature and horizon probes via their shared vacuum correlations \cite{HariPRD2024,MayankPRD2025}, but here the emphasis is on temporal structure and a fully multi-time operational distance.

\paragraph*{Certified comb-level lower bound.}
By Prop.~\ref{prop:LB-choi-comb}, the stitched negativity provides a certified lower bound on the global measure:
\begin{equation}
N(\Upsilon) \ge \frac{2}{d} N_\chi\bigl(\Phi_{k_2:j_1}^{(S_2 \triangleleft S_1)}\bigr) \ge \Delta_{\mathrm{act}}.
\label{eq:activation_LB}
\end{equation}

The stitching corresponds to a valid tester $T_{k_2:j_1}$ that isolates the composite map while idling elsewhere; applying the proposition to this tester yields the inequality. Thus, even if individual segments are CP-divisible (zero negativity), multi-time control certifies global non-Markovianity. This bound is covariant under diamond-preserving reparametrizations and diffeomorphisms (Prop.~\ref{prop:covariance}).

\paragraph*{Microscopic mechanism (weak-coupling picture).}
In the UDW model [Eq.(\ref{eq:Hint})], the Born expansion for the stitched Choi operator yields
\begin{equation}
\chi_{\Phi_{k_2:j_1}^{(S_2 \triangleleft S_1)}} \approx \mathbb{I} + \sum_{m \in S_2, l \in S_1} \eta_{m-l} (\sigma_x \otimes \sigma_x) + O(g^4),
\label{eq:stitched_choi}
\end{equation}
\begin{equation}
    \eta_{m-l} = g^2 \int_{\gamma_m} \int_{\gamma_l} \chi(\lambda) \chi(\lambda') \operatorname{Re} W\bigl(x(\lambda), x(\lambda')\bigr) d\lambda\, d\lambda',
\end{equation}
where cross-segment kernels $\{\eta_{m-l}\}_{l \in S_1, m \in S_2}$ add coherently. Curvature-induced long-range tails in $W$ (e.g., near horizons) can make intra-segment sums positive (CP maps) while cross terms drive overall negativity, yielding $\Delta_{\mathrm{act}} > 0$ and a finite LB (Eq.~\eqref{eq:activation_LB}) \cite{Radzikowski1996,HW2001,Unruh1976,LoukoSatz2008,PKMM2015}.

\paragraph*{Two-probe superactivation (UDW$_A$ + UDW$_B$).}
For two detectors $A$ and $B$ along timelike-separated worldlines interacting with the same field, let $\Upsilon^{AB}$ be the joint process tensor, with marginals $\Upsilon^A$ and $\Upsilon^B$ obtained by idling the other probe. Even if $N(\Upsilon^A) = N(\Upsilon^B) = 0$, cross-probe kernels $\eta^{AB}_{m-l}$ (m on A, l on B) can produce a stitched intermediate $\Phi^{AB}_{k_2:j_1}$ with
\begin{equation}
N(\Upsilon^{AB}) \ge \frac{2}{d} N_\chi\bigl(\Phi^{AB}_{k_2:j_1}\bigr) \ge \Delta_{\mathrm{act}}^{AB} > 0,
\end{equation}
capturing genuine multi-probe effects, invisible in single-detector marginals but emergent from field-mediated correlations in curved spacetime \cite{PKMM2015}.

Since the strategy norm is the operational optimum over causal testers, any positive $\Delta_{\mathrm{act}}$ implies a strictly positive advantage in at least one multi-time discrimination task (e.g., binary hypothesis testing between $\Upsilon$ and its nearest CP-divisible approximation) using an entanglement-assisted, adaptive protocol \cite{Watrous2009}. In Sec.~IV we therefore plot $\Delta_{\mathrm{act}}$ alongside the certified LB/UB bracketing of $N(\Upsilon)$, and we report its dependence on geometric control parameters (e.g., separation, acceleration, and diamond size) to expose which spacetime features most strongly superactivate memory \cite{SmithYardScience2008}.

\section{Memory kernel structures in curved spacetimes: numerical benchmarks and physical insights}
\label{sec:benchmark}

\begin{figure*}[ht]
\centerline{\includegraphics[width=0.9\textwidth]{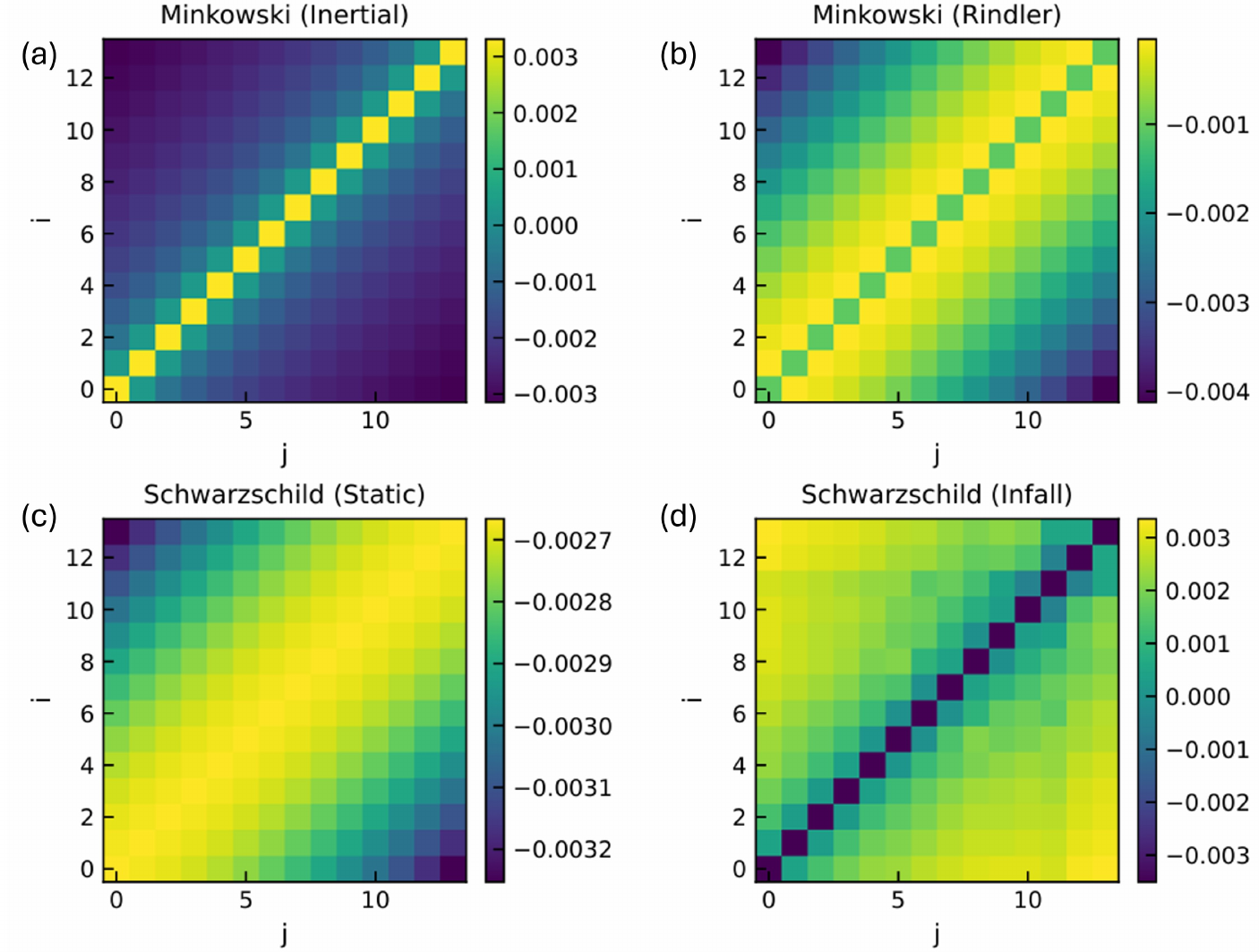}} 
\caption{Heatmaps of the real part of the memory kernel $\eta$-matrix across the four benchmark geometries: (a) Minkowski inertial, (b) Rindler ($a=1.0$), (c) Schwarzschild static ($r=6.0M$), and (d) Schwarzschild infalling (from $r_0=8.0M$). The off-diagonal elements ($\eta_{i-j}$, $i \neq j$) signify non-Markovian memory. These elements are negligible for the inertial case but are significantly enhanced by acceleration (Rindler) and spacetime curvature (Schwarzschild), indicating strong history-dependent dynamics.}
\label{fig:heatmaps}
\end{figure*}

In the preceding sections, we have developed a covariant framework to quantify non-Markovianity within the context of curved spacetime. In this section, we present numerical benchmarks of the covariant non-Markovianity measures introduced in Section \ref{sec:PT_curved} and \ref{sec:measure}, focusing on the memory kernel $\eta$-matrix derived from the process tensor formalism. Our analysis is conducted within a simplified 1+1-dimensional framework, which captures essential features of quantum field correlations in curved spacetimes while allowing for tractable computations~\cite{Birrell1982,Tjoa2022}. The physical system under consideration is an Unruh-DeWitt (UDW) detector, modeled as a two-level quantum system weakly coupled to a massless scalar field $\phi$ along a timelike worldline parameterized by proper time $\tau$. The worldline is partitioned into \(n\) equal slots and each slot is smeared by a smooth, compactly supported switching window (Gaussian/raised–cosine; unit \(L^2\) norm). Unless stated otherwise, the accelerated case is discretized in \emph{rapidity} \(\theta=a\tau\) (the stationary pullback of \(W\) in \(\Delta\theta\)), while inertial/Schwarzschild cases are discretized in proper time. For each pair of slots, we evaluate
\[
\eta_{i-j}\;=\;g^2\!\int_{\gamma_i}\!\!\!\!\mathrm{d}\lambda
\int_{\gamma_j}\!\!\!\!\mathrm{d}\lambda'\,
\chi(\lambda)\chi(\lambda')\,\mathrm{Re}\,W\!\big(x(\lambda),x(\lambda')\big),
\]
This kernel captures pairwise non-Markovian effects: diagonal elements \( \eta_{i-i} \) govern local dissipation and decoherence, while off-diagonals \( \eta_{i-j} \) (for \( i \neq j \)) signal memory backflow mediated by spacetime curvature.

To isolate curvature-induced memory, we employ Hadamard subtraction~\cite{HollandsWald2015PhysRep, DecaniniFolacci2008,Hodkinson2012}: the universal singular part of \( W \) (arising from the Minkowski-like local structure near coincidence) is removed, leaving the regular, state-dependent remainder that encodes global correlations. This subtraction ensures that \( \eta \) reflects genuine non-local effects rather than ultraviolet divergences. Numerically, we discretize the worldline into \( n = 14 \) slots over a total affine parameter range \( L = 12.0 \), with Gaussian switches of width $w = 0.4\,\Delta$ (where $\Delta = L/n$) and use Simpson quadrature with 500 points per integral for accuracy. Geodesic integration for infalling trajectories is handled via \texttt{solve\_ivp} with adaptive Runge-Kutta (RK45) stepping, achieving relative tolerances of \( 10^{-8} \).

We avoid a full comb semidefinite programming (SDP) formulation for the process tensor due to computational complexity: the Choi operator for \( n \) slots scales as \( d^{2n} \) (with \( d = 2 \) for a qubit), rendering SDP optimization over Markovian combs infeasible for \( n \gtrsim 10 \) without specialized hardware~\cite{Watrous2009}. Instead, we benchmark the $\eta$-matrix directly, as it approximates the leading-order non-Markovian kernel in the weak-coupling Dyson expansion~\cite{PollockPRA2018, HuPazZhang1992}. This choice is justified by the perturbative regime (\( g \ll 1 \)), where higher-order terms are suppressed, and aligns with prior studies showing that pairwise kernels capture dominant memory effects in UDW models~\cite{MilzPRXQ2021, BFV2003}.

We examine four paradigmatic geometries: (i) Minkowski inertial, (ii) Rindler accelerated (uniform acceleration \( a \)), (iii) Schwarzschild static (fixed radial coordinate \( r \)), and (iv) Schwarzschild infalling (from initial \( r_0 \)). These span flat-to-curved transitions, allowing direct comparisons. The black hole mass is fixed at \( M = 1 \) (in geometric units), with default parameters \( a = 1.0 \), \( r = 6.0M \), and \( r_0 = 8.0M \).

Figure \ref{fig:heatmaps} plots the real part of the kernel \(\eta_{i-j}\) as a function of indices \(i,j\) for the four geometries. The diagonal ridge reflects short-range self-correlations regulated by the window; the physics resides in how quickly the off-diagonal mass decays and in the \emph{sign structure} away from the diagonal.

\begin{figure*}[ht]
\centerline{\includegraphics[width=0.9\textwidth]{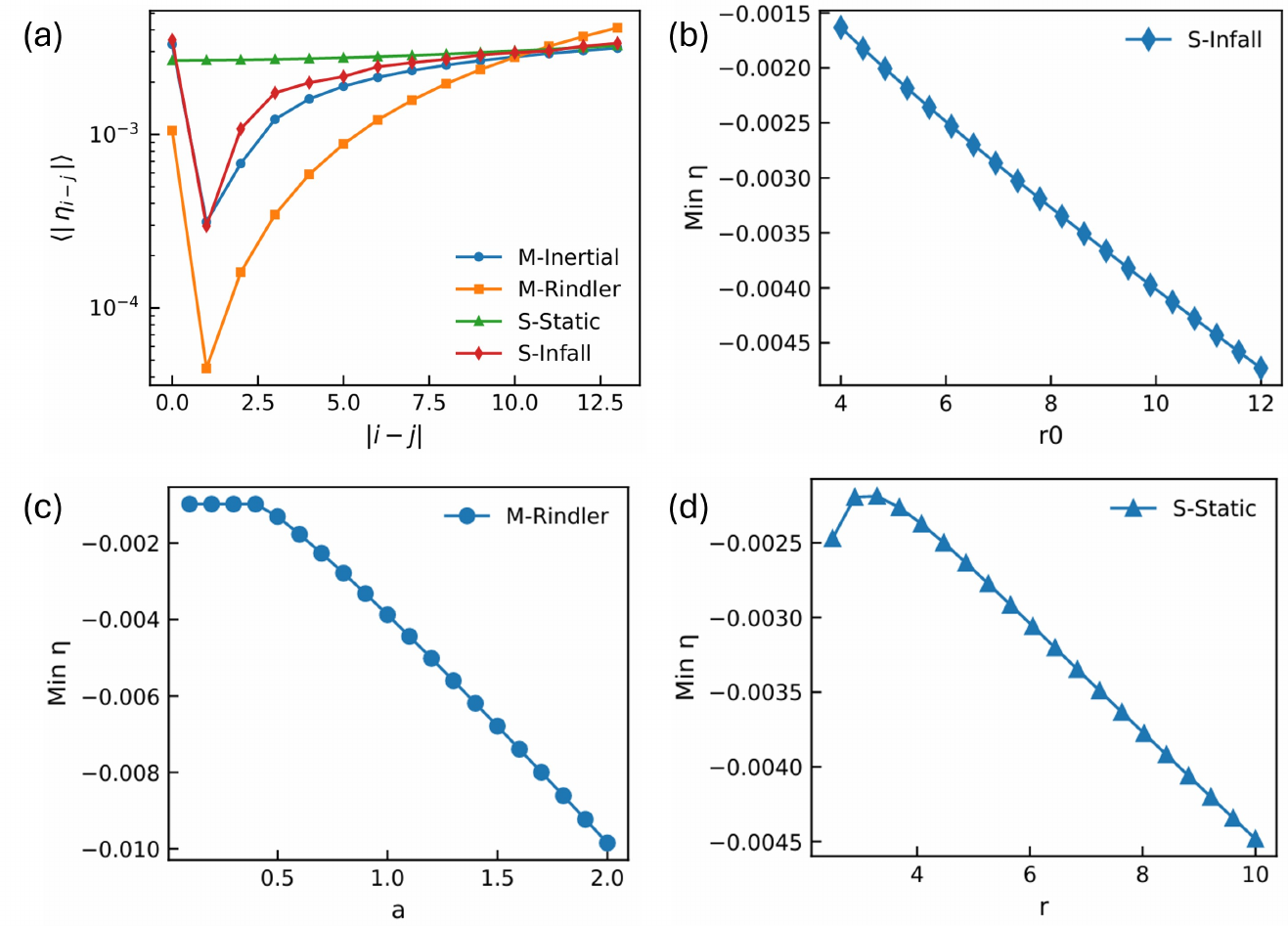}}
\caption{Line cuts and parameter scans of the memory kernel $\eta$‑matrix. (a) Median magnitude $(\langle|\eta_{i-j}|\rangle)$ as a function of slot separation for four worldlines: Minkowski inertial (blue), Rindler accelerated (orange), Schwarzschild static (green), and Schwarzschild infall (red). (b–d) Most negative off‑diagonal entries as functions of initial radius $r_0$ (infall), acceleration a (Rindler), and radial position r (static). Together these plots demonstrate how acceleration and curvature modulate long‑range anticorrelations and memory backflow.}
\label{fig:linecuts}
\end{figure*}

In flat spacetime (1+1D) with an inertial detector, the pulled back Wightman function is
\begin{equation}
  W(s) = -\frac{1}{2\pi} \log |s| + C,
  \label{inertial_W}
\end{equation}

where $C$ incorporates an infrared cutoff, and $s$ represents the proper time separation~\cite{LoukoSatz2008, Birrell1982}. The logarithmic form arises from the infrared-enhanced correlations inherent to lower-dimensional quantum fields, leading to a slower decay than in higher dimensions. Upon double integration against Gaussian switching functions, this yields $\eta_{i-j}$ with positive diagonal entries (from the positive divergence near $s=0$) that transition to negative off-diagonals at moderate $|i-j| \gtrsim 2$ (Figure \ref{fig:heatmaps}(a)). This sign structure reflects non-Markovianity sourced purely by the field's vacuum fluctuations, without curvature or acceleration: short-range positive correlations enhance local dissipation, while long-range negative tails enable partial recoherence over extended times.

For an accelerated (Rindler) trajectory in flat spacetime (Figure \ref{fig:heatmaps}(b)), the Wightman function becomes stationary in proper time, 

\begin{equation}
    W(s) = -\frac{1}{2\pi} \log \left| \sinh \left( \frac{a s}{2} \right) \right| + C',
    \label{rindler_W}
\end{equation}
mirroring a thermal bath at Unruh temperature $T_U = a/(2\pi)$~\cite{Takagi1986}. The exponential decay in the tail (from $\sinh(as/2) \approx e^{a|s|/2}/2$ for large $s$) produces broader negative off-diagonals than the inertial case, with "thicker" blue bands indicating stronger, longer-ranged anticorrelations. Mathematically, this stems from the hyperbolic mapping of coordinates, which periodically extends the correlator in imaginary time, imprinting thermal periodicity. Physically, the acceleration mixes positive and negative frequency modes, yielding Unruh radiation that amplifies memory effects, pushing the dynamics further from Markovianity as quantified by larger off-diagonal magnitudes.

In the static Schwarzschild case (Figure \ref{fig:heatmaps}(c)), the detector at fixed radius $r$ experiences a hybrid of thermal and geometric effects. The Tolman redshift elevates the local temperature to 
\[T(r) = \frac{T_H}{\sqrt{1 - 2M/r}},\]
where $T_H = 1/(8\pi M)$ is the Hawking temperature at infinity~\cite{Candelas1980}. This "warms" the local Wightman function, akin to increasing the effective acceleration in Rindler, resulting in longer-ranged negative tails in $\mathrm{Re}\, W(s)$ and thus thicker off-diagonal bands in $\eta_{i-j}$. However, spacetime curvature modulates this via the Hadamard form's biscalars: the Van Vleck determinant (encoding geodesic focusing) and DeWitt coefficients introduce gentle sign oscillations in $\mathrm{Re}\, W(s)$ at intermediate separations~\cite{HollandsWald2015PhysRep, DecaniniFolacci2008}. These arise from null geodesic convergence near the horizon, tempering the purely negative thermal tail with partial cancellations. The net kernel exhibits broader off-diagonals than inertial but less uniform negativity than Rindler, reflecting a balance where thermal amplification extends correlations while curvature-induced phases soften their coherence.

During radial infall (Figure \ref{fig:heatmaps}(d)), non-stationary motion introduces Doppler shifts and evolving curvature scales, reshaping the kernel's sign texture~\cite{Hodkinson2012}. As the detector accelerates toward the horizon, its velocity relative to field modes induces time-dependent red/blueshifts, mixing frequency components in $\mathrm{Re}\, W(\lambda - \lambda'; v(\lambda), v(\lambda'))$ and driving oscillations. Curvature drift along the trajectory further imprints phases via changing Hadamard coefficients, competing with near-horizon redshift amplification. Near the diagonal, local damping yields negative correlations; at intermediate $|i-j|$, Doppler-curvature interplay flips signs positive (warm bands); far off, horizon-enhanced tails revert to negative. This alternating pattern, negative near, positive mid, negative edge is a spectral fingerprint of non-stationarity, absent in stationary cases. Mathematically, it emerges from the geodesic-integrated pullback, with asymmetry under $i \leftrightarrow j$ due to infall history. Physically, it signals multi-time memory resources: fixed-time marginals (e.g., BLP witnesses) underestimate non-Markovianity by averaging phases, while phase-aligned process-tensor probes can exploit bands for stronger certification via Choi-negativity or diamond-norm distances to Markovian combs.

Figure~\ref{fig:linecuts} displays line cuts of the memory kernel against slot separation together with parameter scans of the kernel’s most negative entry, which we use as a proxy for long–range anticorrelations. Panel~\ref{fig:linecuts}(a) plots the median magnitude $(\langle|\eta_{i-j}|\rangle)$ versus $(|i-j|)$ on semi–log axes for four worldlines: Minkowski inertial (blue), Minkowski Rindler (orange), Schwarzschild static (green), and Schwarzschild infall (red). Across all scenarios except Schwarzschild–static, we observe a dip at $|i-j|=1$, which arises because adjacent slots correspond to partially overlapping causal diamonds: the convolution of the Wightman kernel with two overlapping window functions produces strong destructive cancellation. As the separation grows, the diamonds cease to overla;, this cancellation rapidly weakens, and the median $\langle|\eta_{i-j}|\rangle$ rebounds toward the asymptotic floor set by the intrinsic tail of the correlator (algebraic for inertial/infall, thermal for Rindler/static). This dip is therefore best viewed as a feature of the discretization rather than a universal short-distance signature of memory. The Schwarzschild–static case lacks this feature because proper-time stationarity with a single Tolman thermal scale places the kernel directly on its exponential slope, so the overlap-induced minimum is washed out.

Relative to inertial, the Rindler curve is uniformly suppressed while retaining a similar overall shape. This follows from the KMS structure of the accelerated vacuum at the Unruh temperature $(T_U=a/(2\pi))$: the correlator decays as \[W_{\rm Rindler}(\Delta\tau)\propto \sinh^{-2}\big(\tfrac{a}{2}(\Delta\tau-i\varepsilon)\big),\] yielding an exponential tail at large $(|\Delta\tau|)$. The exponential cutoff reduces long–lag contributions, and KMS antisymmetry near $(\Delta\tau=0)$ enhances adjacent–slot cancellations, producing the characteristic dip at $(|i-j|=1)$.

The Schwarzschild static profile is nearly linear on these semi–log axes—an exponential falloff—without a pronounced $(|i-j|=1)$ dip. This is the signature of proper–time stationarity at fixed radius, where a single Tolman–local thermal scale
\[
\xi(r)\sim \frac{1}{\pi T_{\rm loc}(r)},\qquad
T_{\rm loc}(r)=\frac{T_H}{\sqrt{1-2M/r}}
\]
sets the decay. Once the slot width exceeds microscopic cutoffs, the kernel sits on this thermal slope and adjacent–slot cancellations are muted by translation symmetry. By contrast, the infalling Schwarzschild curve tracks the inertial one at small $(|i-j|)$, consistent with local equivalence: over short separations a freely–falling detector samples near–vacuum correlations. At intermediate and long lags, the accumulated gravitational redshift chirps the effective sampling frequency and slightly lifts the curve relative to inertial, evidencing weak non–stationarity.

Figure~\ref{fig:linecuts}(b)–(d) quantify how geometry and kinematics tune the most negative off–diagonal entry. For infall, $(\min\eta)$ becomes more negative as the initial radius $(r_0)$ increases [Figure~\ref{fig:linecuts}(b)], indicating that longer exposure to gentle curvature gradients amplifies the negative lobe at intermediate lags. In the Rindler case, $(\min\eta)$ plateaus for small accelerations and then decreases approximately linearly beyond a knee at $(a\Delta\tau=\mathcal O(1))$ [Figure~\ref{fig:linecuts}(c)]: below the knee the Unruh correlation time $(\xi_U\propto 1/a)$ exceeds the slot width and the kernel is effectively inertial; once resolved, thermal anticorrelations deepen with (a). For static Schwarzschild, $(\min\eta)$ shows a shallow shoulder at small radii before a gentle outward decrease [Figure~\ref{fig:linecuts}(d)]. This shape reflects the competition between Tolman heating, which shortens $(\xi(r))$ and strengthens anticorrelations near the horizon, and redshift of the proper-time step $(\Delta\tau=\sqrt{1-2M/r},\Delta t)$ at fixed coordinate sampling, which reduces adjacent–slot overlap and partially offsets thermal screening. Far from the horizon the local temperature drops, the correlation length grows, and $(\min\eta)$ relaxes toward the cooler, nearly flat–space limit.

\begin{figure*}[ht]
\centerline{\includegraphics[width=\textwidth]{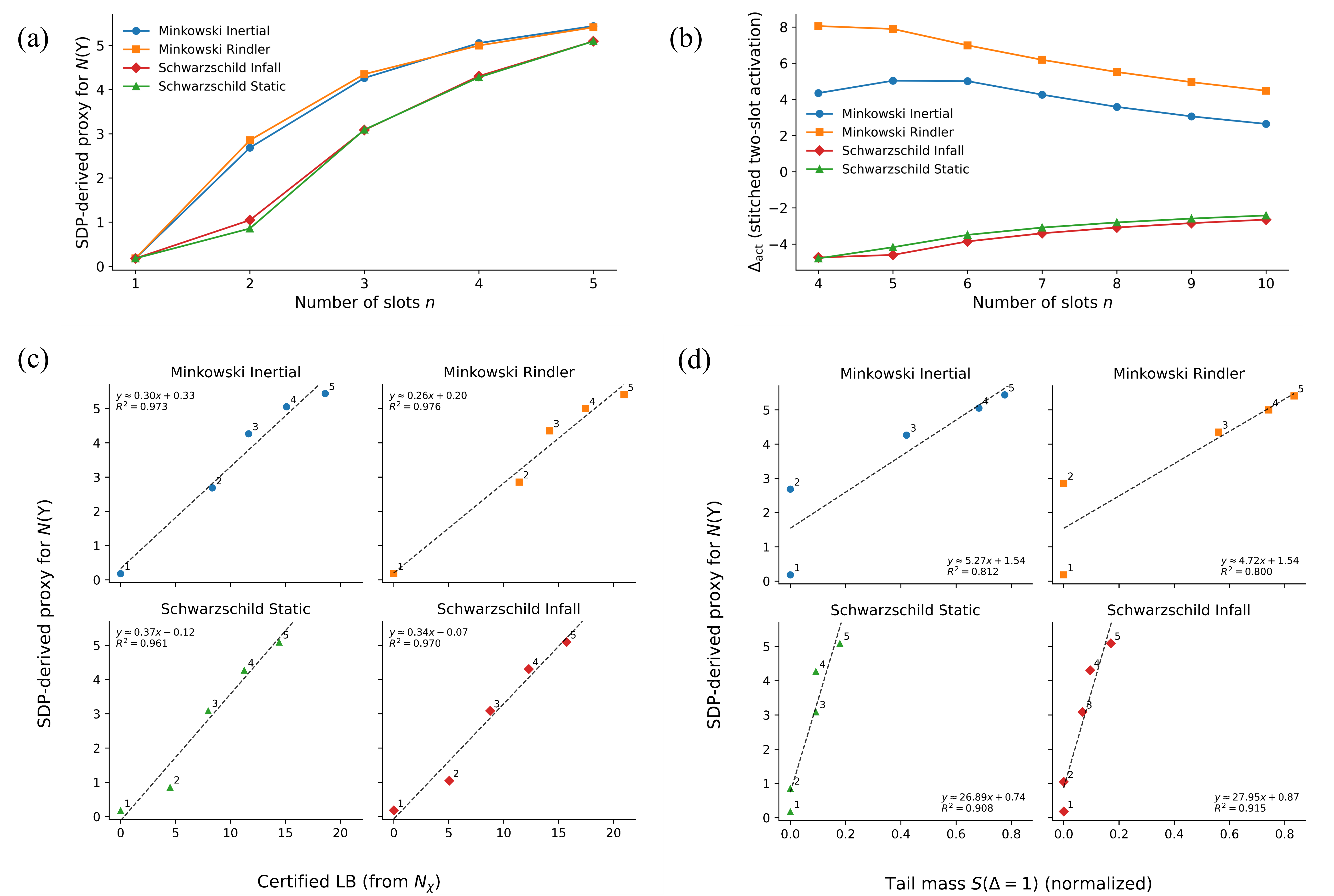}}
\caption{(Color online) Operational bounds and SDP-based estimates of covariant non-Markovianity across the four benchmark worldlines (Minkowski inertial, Minkowski Rindler, Schwarzschild static, Schwarzschild infall). 
(a) SDP-derived lower-bound proxy for $N(\Upsilon)$, given by the summed absolute prediction error $\sum_{\text{testers}}|\Delta|$, as a function of the number of slots $n$, showing a clear enhancement of non-Markovianity for accelerated and infall trajectories relative to the inertial case. 
(b) Stitched two-slot activation $\Delta_{\text{act}}$ versus $n$, obtained from the excess negativity of a nonlocal two-slot block over its constituent single-slot blocks~\cite{Watrous2009, FawziRennerCMP2015}, which reproduces the same geometry ordering and provides a complementary witness of operational memory activation. 
(c) Geometry-resolved scatter plots of the SDP proxy for $N(\Upsilon)$ versus the certified lower bound from intermediate Choi negativities, together with linear fits and $R^2$ values, demonstrating that the Choi-based bound tracks the full operational distance with nearly linear behavior in each geometry. 
(d) Corresponding scatter of the SDP proxy versus the normalized kernel tail mass $S(\Delta=1)$, indicating that processes with heavier long-range tails in $\eta_{i-j}$ systematically display larger non-Markovianity while preserving the hierarchy observed in panels~(a) and (b).}
\label{fig:limits}
\end{figure*}

Figure~\ref{fig:limits} gathers these kernel-level observations into fully operational statements about the process tensor. Figure~\ref{fig:limits}(a) shows, for $n \leq 5$, the SDP-derived lower-bound proxy for $N(\Upsilon)$ obtained by fitting the tester data to the nearest CP-divisible product of single-step channels (see \ref{app:SDP}. For $n \leq 5$ we have performed the full SDP on a single workstation; however, the exponential growth of the underlying Hilbert-space dimension makes such optimizations rapidly intractable for larger $n$. The bound is quantified by the summed absolute prediction error $\sum_{\text{testers}}|\Delta|$, which lower-bounds the strategy-norm distance between the true process tensor and its best Markovian approximation within this restricted class. For our benchmark parameters, the Minkowski inertial worldline yields the smallest values across all $n$, while uniform acceleration in Minkowski and motion in the Schwarzschild background lead to visibly larger deviations from Markovianity. Within each spacetime family, the Rindler trajectory lies above the inertial one, and the infalling worldline develops stronger deviations than the static Schwarzschild detector. Comparisons between the Minkowski and Schwarzschild families should be read as parameter dependent: different choices of $a$, $M$, or $r_0$ would shift the relative positions of the curves, even though the basic trend of “more structure in the correlator $\Rightarrow$ larger $N(\Upsilon)$’’ remains intact.

Figure~\ref{fig:limits}(b) reports the stitched two-slot activation $\Delta_{\text{act}}$ as a function of $n$, constructed from the excess Choi negativity of a composite two-slot block over its constituent single-slot intermediates. By design, $\Delta_{\text{act}}$ isolates superactivation-type effects: it vanishes whenever non-Markovianity is already witnessed by individual intermediates, and becomes positive only when stitching operations across well-separated diamonds unlocks additional memory resources. For the trajectories considered here, $\Delta_{\text{act}} \geq 0$ in all cases and follows the same qualitative pattern as figure~\ref{fig:limits}(a): the inertial process is closest to Markovian, while accelerated and black-hole worldlines exhibit larger activation within their respective families. The initial growth and subsequent saturation of $\Delta_{\text{act}}$ with $n$ indicate that the long-range kernel contributions are converted into operational advantage up to the point where additional slots cease to probe genuinely new correlations.

The scatter plots in Figure~\ref{fig:limits}(c) make the link between intermediate maps and the full process explicit. Here, the SDP proxy for $N(\Upsilon)$ is plotted against the certified lower bound $\max_{j<k}\tfrac{2}{d}N_\chi(\Phi_{k{:}j})$ obtained from the most non-CP-divisible intermediate for each geometry and $n$. The points in each panel cluster around an approximately linear trend, with best-fit lines and $R^2$ values close to unity. This shows that, in the weak-coupling and few-slot regime explored here, the negativity of a single “most correlated’’ intermediate map already captures most of the operational distance to the Markovian set. In practice, an experiment that can tomographically reconstruct only selected intermediates, rather than the full comb, still obtains a robust predictor for $N(\Upsilon)$ that is stable under changes of trajectory within a given spacetime.

Figure~\ref{fig:limits}(d) replaces this Choi-level quantity by a purely kernel-level measure, the normalized tail mass $S(\Delta)$ obtained by summing $|\eta_{i-j}|$ over separations $|i-j|\geq \Delta$ (here $\Delta=1$). The roughly monotone, weakly nonlinear relationship between the SDP proxy and $S(\Delta)$ across all trajectories confirms a simple “kernel-mass $\to$ operational-distance’’ principle: once the universal short-distance singularity has been subtracted, processes with heavier long-range tails in the regular part of the Wightman kernel tend to exhibit larger $N(\Upsilon)$. Within each family (Minkowski or Schwarzschild), this tail mass tracks the same ordering that appears in the SDP-derived bounds and in $\Delta_{\text{act}}$, indicating that the pairwise kernels distilled in Figure~\ref{fig:heatmaps} and Figure~\ref{fig:linecuts} already contain the essential information needed to anticipate the size of the covariant non-Markovian resource.

Although our explicit numerics are carried out in an effective $(1+1)$-dimensional setting, the qualitative hierarchy between trajectories is expected to persist in $(3+1)$ dimensions with a smaller overall scale of the memory effects. In $(3+1)$-dimensional Minkowski spacetime the vacuum Wightman function along a timelike worldline decays as $1/s^{2}$, whereas in $(1+1)$ dimensions it exhibits a logarithmic dependence (Eq.~\ref{inertial_W}). After smearing with smooth windows of fixed macroscopic size, this faster decay reduces the absolute magnitude of the off-diagonal kernel entries $\eta_{i-j}$ and hence the tail mass $S(\Delta)$, but does not alter the ordering between inertial, accelerated, and near-horizon trajectories. Acceleration and the presence of a Schwarzschild horizon still imprint KMS-like periodicity and near-horizon enhancements in the pulled-back correlator, which translate into parametrically longer-lived kernel tails than in flat inertial motion. Since both the Choi-negativity lower bound and the SDP-based proxy for $N(\Upsilon)$ depend linearly on these tails, the operational advantage in $(3+1)$ dimensions is suppressed only by an order-one factor set by microscopic versus slot scales, rather than exponentially in the number of dimensions. Analytical estimates for the $(3+1)$-dimensional kernels for the same trajectories, presented in the appendices, support this expectation.

\section{Conclusion}
\label{sec:conclusion}

In this work we have introduced a fully covariant, operational framework for quantifying quantum non-Markovianity along arbitrary timelike worldlines in curved spacetimes. By constructing process tensors from causal diamonds and defining non-Markovianity via the strategy-norm distance to the set of CP-divisible combs, we obtained a resource monotone that is manifestly invariant under reparametrizations and causal-structure-preserving diffeomorphisms. Memory thereby emerges as an intrinsic geometric feature of spacetime correlations rather than an artifact of foliation or coordinate choice.

Our analysis reveals that acceleration and spacetime curvature act as genuine resources for quantum memory. Uniform acceleration in flat spacetime and the presence of horizons in Schwarzschild geometry induce pronounced long-range temporal correlations that translate into a quantifiable operational advantage in multi-time processing tasks. Notably, these effects can superactivate memory in multi-probe settings and persist even when local pairwise diagnostics suggest near-Markovian behavior. The framework thus provides relativistic quantum information protocols ranging from quantum clocks and metrology near horizons to communication with accelerated or gravitating parties, with rigorous tools to determine when curvature and horizons enhance performance and when they must be mitigated as noise.

Several natural extensions present themselves. Generalizing the present results to (3+1) dimensions, to spinning or gauge fields, and to dynamical spacetimes (gravitational collapse, cosmology) will reveal characteristic memory signatures of time-dependent horizons and particle creation. Embedding the measure into a full resource theory of covariant non-Markovianity, designing curvature-aware error correction and feedback protocols, and exploring multi-detector networks with entangled probes are immediate information-theoretic directions. Finally, combining the perturbative process-tensor construction with tensor-network or other nonperturbative methods promises access to strong-coupling and quantum-gravity regimes in which spacetime itself participates dynamically in the memory structure.

With these tools, quantum memory joins the catalog of relativistic phenomena alongside the Unruh effect and Hawking radiation that blur the boundary between quantum information and geometry, suggesting that the causal structure of spacetime is not merely a stage but an active resource in quantum processing.

\section*{DATA AVAILABILITY}

The numerical data and analysis scripts that support the findings of this work are openly available in the Zenodo repository \cite{WaghmareZenodo2025}.

\appendix
\setcounter{section}{0}
\renewcommand{\thesection}{\Alph{section}}

\section{Pullback of the Wightman function to the Rindler trajectory}
\label{app:rindler_W}

We start from the Minkowski Wightman function for a massless scalar field in $(3{+}1)$ dimensions in the vacuum state,
\begin{align}
G^+(x,x') 
&= \langle 0|\phi(x)\phi(x')|0\rangle \notag\\
&= \frac{1}{4\pi^2\big[(t-t'-i\epsilon)^2 - |\mathbf{x}-\mathbf{x}'|^2\big]} ,
\end{align}
with the usual $i\epsilon$ prescription.

We parameterize the uniformly accelerated worldline by the rapidity $\theta$ with $\theta = a\tau$, and choose
\begin{equation}
t(\theta)=\frac{1}{a}\sinh\theta, \qquad 
x(\theta)=\frac{1}{a}(\cosh\theta-1), \qquad 
y=z=0 .
\end{equation}
The overall spatial shift in $x$ has no physical effect on the Wightman function.

The invariant interval between two points on the trajectory is
\begin{align}
\Delta s^2 
&\equiv \big(t(\theta)-t(\theta')-i\epsilon\big)^2 
      - \big|\vec x(\theta)-\vec x(\theta')\big|^2 \notag\\[2pt]
&= a^{-2}\Big[\sinh\theta-\sinh\theta' - i a\epsilon\Big]^2 \notag\\
&\quad - a^{-2}\Big[\cosh\theta-\cosh\theta'\Big]^2 \notag\\[2pt]
&= -\,4 a^{-2}\sinh^2\!\left(\frac{\Delta\theta - i\epsilon}{2}\right),
\end{align}
where $\Delta\theta = \theta-\theta'$ and hyperbolic identities have been used in the last step. Substituting into $G^+$ yields the Rindler pullback
\begin{equation}
W(\Delta\theta) 
\equiv G^+\big(x(\theta),x(\theta')\big) 
= -\frac{a^2}{16\pi^2\sinh^2\!\left[\frac{\Delta\theta - i\epsilon}{2}\right]}.
\end{equation}
 The overall minus sign originates from the negative invariant interval along the timelike trajectory; the $i\epsilon$ prescription selects the standard retarded/advanced analytic continuation.

\paragraph*{Large-separation asymptotic.}
For large rapidity separations $\Delta\theta\gg1$ we have
\begin{equation}
\sinh\!\left(\frac{\Delta\theta}{2}\right)\simeq \frac{1}{2}e^{\Delta\theta/2},
\end{equation}
so that
\begin{equation}
W(\Delta\theta) \simeq -\frac{a^2}{4\pi^2}\,e^{-\Delta\theta}, 
\qquad \Delta\theta\gg1.
\end{equation}
This exponential decay provides the underlying origin of the horizon-induced long-range memory discussed in the main text: correlations between widely separated slots in rapidity are suppressed only exponentially, rather than in a power-law fashion.

\section{Evaluation of \texorpdfstring{$\eta_{i-j}$}{η\_{i-j}} for smooth switching}
\label{app:perturb}

We recall the definition of the memory kernel elements
\begin{equation}
\eta_{i-j} 
= g^2 \int_{\gamma_i} d\theta 
    \int_{\gamma_j} d\theta' \;
    \chi_\sigma(\theta)\,\chi_\sigma(\theta')\,
    \Re W(\theta-\theta'),
\end{equation}
where $\chi_\sigma$ is a normalized smooth switching function supported (or effectively supported) on a segment of width $\sim \Delta\theta$, and $W$ is the pullback of the Wightman function to the trajectory. 

For analytic control we choose a normalized Gaussian switching profile,
\begin{equation}
\chi_{w}(\theta) 
= \frac{1}{(2\pi w^2)^{1/4}}
  \exp\!\left[-\frac{(\theta-\theta_k)^2}{4 w^2}\right],
\end{equation}
centered at the midpoint $\theta_k$ of segment $\gamma_k$, with normalization $\int d\theta\,\chi_{w}^2(\theta)=1$.

For well-separated segments $|i-j|$ with $|i-j|\Delta\theta \gg \sigma$, we may use the large-separation asymptotic of the Rindler Wightman function,
\begin{equation}
\Re W(\Delta\theta) \simeq -\frac{a^2}{4\pi^2} e^{-|\Delta\theta|}.
\end{equation}
For definiteness, take $i>j$ so that $\Delta\theta_{ij}=(i-j)\Delta\theta>0$. Then
\begin{align}
\eta_{i-j} 
&\simeq -\frac{g^2 a^2}{4\pi^2}
   \int d\theta \int d\theta' \;
   \chi_{w}(\theta)\,\chi_{w}(\theta')\, e^{-(\theta-\theta')} \notag\\
&\simeq -\frac{g^2 a^2}{4\pi^2}\,
   e^{-\Delta\theta_{ij}}
   \Bigg[\int d\theta\,\chi_{w}(\theta)\,e^{-\theta}\Bigg]
   \Bigg[\int d\theta'\,\chi_{w}(\theta')\,e^{\theta'}\Bigg],
\end{align}
where in the second line we have expanded around the midpoints of the two windows and used the fact that the separation between midpoints is $\Delta\theta_{ij}$. For windows symmetric about their respective midpoints and of width $w\lesssim\Delta\theta$, the integrals are dominated by $\theta\approx\theta_i$ and $\theta'\approx\theta_j$ and scale as
\begin{equation}
\int d\theta\,\chi_{w}(\theta)\,e^{-\theta} \sim e^{-\theta_i}w, 
\qquad
\int d\theta'\,\chi_{w}(\theta')\,e^{\theta'} \sim e^{\theta_j}w,
\end{equation}
up to $O(1)$ numerical factors. Their product therefore contributes an additional factor $e^{-\Delta\theta_{ij}}\sigma^2$, leading to
\begin{equation}
\eta_{i-j} 
\simeq -C_\eta\, g^2 a^2 \sigma^2\, e^{-2\Delta\theta_{ij}},
\qquad \Delta\theta_{ij}=(i-j)\Delta\theta,
\end{equation}
with $C_\eta=O(1)$ determined by the precise window shape. Thus, for smooth segment-sized windows ($w\sim\Delta\theta$), the off-diagonal kernel entries decay exponentially with slot separation and scale quadratically with the effective segment width, in agreement with the qualitative behaviour used in the main text.

A closed-form expression for $\eta_{i-j}$ with full Gaussian factors and the exact Rindler Wightman function can be obtained by standard Gaussian integration. Its explicit form is lengthy and not needed for the qualitative scaling arguments presented in Sec.~\ref{sec:benchmark}.

\paragraph*{Remarks on regularization and UV behaviour.}
Sharp (rectangular) switching, $\chi(\theta)=\Delta\theta^{-1/2}$ on each segment, leads to spurious high-frequency contributions and distributional boundary terms at the switching edges, as is well known in the Unruh-DeWitt literature (see, e.g., Ref.~\cite{LoukoSatz2008}). For physical predictions we therefore advocate using a smooth family $\chi_\sigma$ and verifying the robustness of the results either in the combined limit $w\to 0$ taken after the continuum limit, or at finite $w$ comparable to the slot size where observables are insensitive to moderate variations in $w$.

\section{Wightman function with transverse velocity}
\label{app:transverse}

We next consider a trajectory with a transverse component in the $y$ direction. The worldline can be written as
\begin{equation}
x(\theta) = \big(t(\theta),x(\theta),y(\theta),0\big), 
\qquad 
y(\theta)=\frac{v}{\sqrt{1-v^2}}\frac{\theta}{a},
\end{equation}
where $v$ is the (constant) transverse velocity in Minkowski coordinates and $t(\theta),x(\theta)$ are as in the Rindler case above. The invariant separation between two points on the trajectory is
\begin{align}
\Delta s^2 
&= \big(t(\theta)-t(\theta')-i\epsilon\big)^2 
   - \big(x(\theta)-x(\theta')\big)^2
   - \big(y(\theta)-y(\theta')\big)^2 .
\end{align}
The longitudinal part reproduces the Rindler expression of Appendix~\ref{app:rindler_W}, while the transverse contribution is
\begin{equation}
\Delta y 
= y(\theta)-y(\theta') 
= \frac{v}{a\sqrt{1-v^2}}\,\Delta\theta .
\end{equation}
Combining these terms and using the hyperbolic identities as before, one finds
\begin{equation}
\Delta s^2 
= -4 a^{-2}\sinh^2\!\left(\frac{\Delta\theta - i\epsilon}{2}\right)
  + \left(\frac{v\,\Delta\theta}{a\sqrt{1-v^2}}\right)^2.
\end{equation}
The corresponding Wightman function pulled back to the trajectory is therefore
\begin{equation}
W(\theta,\theta') 
= \frac{1}{4\pi^2\left[\big(\tfrac{v\,\Delta\theta}{a\sqrt{1-v^2}}\big)^2 
    - 4\sinh^2\!\left(\tfrac{\Delta\theta-i\epsilon}{2}\right)\right]}.
\end{equation}

The second term in the denominator is inherited from the Rindler motion, whereas the first arises from the transverse drift. Their relative magnitude controls whether the denominator can become small for certain rapidity separations, potentially enhancing correlations at finite $\Delta\theta$. For general time-dependent transverse motion $y(\theta)$, stationarity in $\Delta\theta$ is lost altogether; the expression above provides the simplest constant-velocity benchmark.

\section{Semidefinite programs and numerical evaluation}
\label{app:SDP}

Let $\Upsilon_{n:1}$ denote the Choi operator of an $n$-slot process tensor acting on the sequence of system input/output Hilbert spaces
$\{\mathcal{H}^{\mathrm{i}}_k,\mathcal{H}^{\mathrm{o}}_k\}_{k=1}^n$
(in the same slot ordering used throughout the main text). For any Hermitian operator
\begin{equation}
\Delta \;:=\; \Upsilon_{n:1} - \widetilde{\Upsilon}_{n:1},
\end{equation}
the comb (strategy) norm $\|\Delta\|_{\mathrm{comb}}$ admits an operational interpretation as the optimal advantage in distinguishing the two processes using an arbitrary causal multi-time tester (co-strategy). Concretely, for two equiprobable hypotheses, the optimal success probability satisfies
\begin{equation}
p_{\mathrm{succ}}^\star \;=\; \frac{1}{2} + \frac{1}{4}\,\|\Delta\|_{\mathrm{comb}}.
\label{eq:psucc_comb}
\end{equation}
Equivalently,
\begin{equation}
\|\Delta\|_{\mathrm{comb}}
\;=\; 2 \max_{\{T_0,T_1\}} \Bigl\{ \mathrm{Tr}\bigl[\Delta\,T_0\bigr] \Bigr\},
\label{eq:comb_as_tester}
\end{equation}
where $\{T_0,T_1\}$ is a two-outcome $n$-slot tester, and $T_0,T_1$ are positive operators on the same tensor-product space as $\Upsilon_{n:1}$ satisfying the standard causal normalization constraints for testers.

A convenient SDP form uses the fact that $T := T_0+T_1$ is a \emph{deterministic tester} (co-strategy). Introducing intermediate normalization operators
$T^{(k)}$ ($k=1,\dots,n$) and an initial state $\rho$ on $\mathcal{H}^{\mathrm{i}}_1$, the optimization \eqref{eq:comb_as_tester} can be written as:
\begin{align}
\textbf{maximize:}\quad
& \mathrm{Tr}\bigl[\Delta\,T_0\bigr]
\\
\textbf{subject to:}\quad
& T_0 \succeq 0, \qquad T^{(n)} \succeq 0, \qquad 0 \preceq T_0 \preceq T^{(n)},
\nonumber\\
& \mathrm{Tr}_{\mathrm{o}_n}\!\left[T^{(n)}\right] = \mathbb{I}_{\mathrm{i}_n}\otimes T^{(n-1)},
\nonumber\\
& \mathrm{Tr}_{\mathrm{o}_{n-1}}\!\left[T^{(n-1)}\right] = \mathbb{I}_{\mathrm{i}_{n-1}}\otimes T^{(n-2)},
\nonumber\\
& \hspace{4.2em}\vdots
\nonumber\\
& \mathrm{Tr}_{\mathrm{o}_2}\!\left[T^{(2)}\right] = \mathbb{I}_{\mathrm{i}_2}\otimes T^{(1)},
\nonumber\\
& \mathrm{Tr}_{\mathrm{o}_1}\!\left[T^{(1)}\right] = \mathbb{I}_{\mathrm{i}_1}\otimes \rho,
\nonumber\\
& \rho \succeq 0,\qquad \mathrm{Tr}[\rho]=1.
\label{eq:comb_norm_SDP}
\end{align}
With the normalization used in \eqref{eq:psucc_comb}, the comb norm is then obtained as
\begin{equation}
\|\Delta\|_{\mathrm{comb}} \;=\; 2\,\mathrm{opt}\bigl(\eqref{eq:comb_norm_SDP}\bigr).
\end{equation}
We use \eqref{eq:comb_norm_SDP} as the exact evaluation of $\|\cdot\|_{\mathrm{comb}}$ for fixed $\Delta$ in the small-$n$ regime where the SDP is tractable.

In practice, we implement the SDP using CVXPY (Python) or CVX (MATLAB) with a reliable interior-point solver such as MOSEK or SCS. For qubit detectors and moderate numbers of slots ($n\lesssim 4$-$5$), the resulting SDPs are numerically tractable; the memory cost scales exponentially with $n$ because the Choi operator of an $n$-slot comb acts on a Hilbert space of dimension $d^{2n}$.

\paragraph*{Practical solver details.}
In numerical implementations it is advantageous to:
\begin{itemize}
\item exploit sparse matrix structures for comb and constraint operators;
\item set solver tolerances (absolute and relative) consistent with the problem scale (e.g., \texttt{abstol} $\sim 10^{-8}$, \texttt{reltol} $\sim 10^{-6}$);
\item verify a posteriori that reconstructed marginals are CPTP within numerical tolerance by checking positivity and trace preservation of the induced maps.
\end{itemize}
The SDP bounds on $N(\Upsilon)$ reported in Sec.~\ref{sec:benchmark} were obtained using this formulation, with additional symmetries (e.g., identical slot structure) exploited whenever applicable.

\section{Tensor-network mapping and convergence}
\label{app:tn}

For completeness we summarize a nonperturbative numerical scheme that can be used to benchmark the perturbative memory-kernel analysis: a chain mapping of the scalar field followed by tensor-network simulation of the resulting open chain.

\paragraph*{Chain mapping.}
The field degrees of freedom coupled to the Unruh--DeWitt detector are mapped to a semi-infinite one-dimensional bosonic chain using standard orthogonal-polynomial techniques. After specifying a spectral density $J(\omega)$ and discretizing the continuum up to a cutoff frequency $\Lambda$, one performs a unitary transformation from the original mode basis to a set of effective chain modes with nearest-neighbour couplings. The Hamiltonian then takes the form
\begin{equation}
H = H_{S} + H_{\text{chain}} + H_{S\text{-chain}},
\end{equation}
where $H_{S}$ describes the detector, $H_{\text{chain}}$ is an approximately nearest-neighbour bosonic chain, and $H_{S\text{-chain}}$ couples the detector locally to the first site of the chain. Local bosonic Hilbert spaces are truncated to $d_{b}$ levels, and the discretization (frequency grid and orthogonal polynomials) is chosen to resolve both low- and high-frequency features of $J(\omega)$ (e.g., logarithmic grids when appropriate).

\paragraph*{MPO simulation and convergence.}
Time evolution of the detector+chain system is performed using standard matrix-product-operator (MPO) or time-evolving block-decimation (TEBD) algorithms with bond dimension $\chi$. Convergence is assessed by increasing $\chi$ and the local cutoff $d_{b}$ until changes in relevant observables (such as the reduced detector state or the operational measure $N(\Upsilon)$ extracted from reconstructed kernels) fall below a preset threshold.

For the weak-coupling benchmarks considered here, representative parameter choices are
\[
\chi \in \{50,100,200\}, \quad d_{b} \in \{4,6,8\}, \quad \Delta t \sim 10^{-2},
\]
with $\Delta t$ measured in units of the detector energy gap. The results quoted in the main text are obtained with $\chi = 100$ and $d_{b} = 6$, and have been spot-checked against $\chi = 200$, $d_{b} = 8$ for selected parameter sets, with discrepancies well below $10^{-3}$ in the quantities of interest. Within the explored regime, the perturbative kernel-based diagnostics and the nonperturbative tensor-network simulations are therefore mutually consistent.

\paragraph*{Extension to stronger coupling.}
Although the present work focuses on the weak-coupling regime, the tensor-network construction is, in principle, nonperturbative in $g$ and can be extended to explore stronger couplings $g \sim 1$, such as those relevant near horizons where large redshifts and enhanced backreaction may render the Born approximation unreliable. In practice, feasibility is controlled by the required chain length and bond dimension needed to capture longer correlation times, potentially necessitating advanced compression strategies or GPU-accelerated implementations. Breakdown of the second-order expansion is expected when higher-order terms in the Dyson series become comparable to the leading contribution, e.g., when
\(
g^{2} \int \mathrm{d}\lambda\,\mathrm{d}\lambda'\, C(\lambda,\lambda') \sim 1,
\)
signalling the onset of significant non-Gaussian field effects and detector–field entanglement. In that regime, reconstructing $\Upsilon$ directly from the tensor-network evolution provides a natural nonperturbative continuation of the covariant framework developed here.

\bibliographystyle{unsrtnat}
\bibliography{references}
\end{document}